\documentclass[onecolumn,a4paper,fleqn,usenatbib]{mnras}
\usepackage{newtxtext,newtxmath}

\usepackage[T1]{fontenc}
\usepackage{ae,aecompl}

\usepackage{graphicx}	
\usepackage{amsmath}	
\usepackage{amssymb}	
\usepackage{mhchem}

\title[Tensile strength of dust aggregates]{The tensile strength of dust aggregates consisting of small elastic grains: constraints on the size of condensates in protoplanetary discs}

\author[H. Kimura et al.]{
Hiroshi Kimura,\thanks{E-mail: hiroshi\_kimura@perc.it-chiba.ac.jp}
Koji Wada,
Fumi Yoshida,
Peng K. Hong,
Hiroki Senshu,
\newauthor
Tomoko Arai,
Takayuki Hirai,
Masanori Kobayashi,
Ko Ishibashi
and Manabu Yamada
\\
Planetary Exploration Research Center (PERC), Chiba Institute of Technology, 
Tsudanuma 2-17-1, Narashino, Chiba 275-0016, Japan\\
}

\date{Accepted 2020 June 4. Received 2020 June 3; in original form 2020 May 2}

\pubyear{2020}

\begin{document}
\volume{496}
\label{firstpage}
\pagerange{1667--1682}
\maketitle

\begin{abstract}
A consensus view on the formation of planetesimals is now exposed to a threat, since recent numerical studies on the mechanical properties of dust aggregates tend to dispute the conceptual picture that submicrometer-sized grains conglomerate into planetesimals in protoplanetary discs.
With the advent of precise laboratory experiments and extensive computer simulations on the interaction between elastic spheres comprising dust aggregates, we revisit a model for the tensile strength of dust aggregates consisting of small elastic grains.
In the framework of contact mechanics and fracture mechanics, we examine outcomes of computer simulations and laboratory experiments on the tensile strength of dust aggregates.
We provide a novel analytical formula that explicitly incorporates the volume effect on the tensile strength, namely, the dependence of tensile strength on the volume of dust aggregates.
We find that our model for the tensile strength of dust aggregates well reproduces results of computer simulations and laboratory experiments, if appropriate values are adopted for the elastic parameters used in the model.
Moreover, the model with dust aggregates of submicrometer-sized grains is in good harmony with the tensile strength of cometary dust and meteoroids derived from astronomical observations.
Therefore, we reaffirm the commonly believed idea that the formation of planetesimals begins with conglomeration of submicrometer-sized grains condensed in protoplanetary discs.
\end{abstract}




\section{Introduction} \label{sec:intro}

The condensation of gas into solid minute grains and the subsequent aggregation of dust grains into planetesimals are believed to be the sequence of events in protoplanetary discs that leads to the formation of planets.
There is a common belief that the constituent grains of dust aggregates in protoplanetary discs have a radius of submicometers, owing to so much evidence of submicrometer-sized grains conglomerated into planetesimals:
A nucleation theory implies that the formation of grains in a protoplanetary disc with solar composition is in accordance with the submicrometer size of condensates in the solar nebula \citep{yamamoto-hasegawa1977};
The spectral and angular dependences of brightness and polarization of cometary comae measured in the visible wavelength range cannot be simultaneously reproduced unless aggregates are composed of submicrometer-sized grains with a radius of $r_0 = 0.1~\micron$ \citep*{kimura-et-al2003,kimura-et-al2006};
Infrared spectral features of forsterite observed in thermal emission from cometary comae are inevitably attributed to porous dust aggregates of constituent grains smaller than micrometers in radius \citep*{okamoto-et-al1994,kolokolova-et-al2007};
The stratospheric collection of interplanetary dust particles (IDPs) by NASA identifies the chondritic porous (CP) subset of IDPs to be of cometary origin and to be aggregates of submicron grains \citep[e.g.][]{brownlee1985};
AFM topographic images of dust aggregates collected in the coma of comet 67P/Churyumov-Gerasimenko (hereafter, 67P/C-G) by MIDAS onboard Rosetta demonstrate without doubt that the constituent grains of dust aggregates in 67P/C-G are submicrometer in radius \citep{bentley-et-al2016,mannel-et-al2016};
The mechanical and electric properties of dust aggregates in the coma of comet 67P/C-G measured by COSIMA/Rosetta are also in harmony with the picture that submicrometer-sized constituent grains with $r_0 = 0.1~\micron$ make up the aggregates \citep{kimura-et-al2020};
\citet{weidenschilling1984,weidenschilling1997} assumed a slightly larger constituent grains with $r_0 = 0.5~\micron$ to model the growth of dust aggregates in protoplanetary discs without justification of the grain size \citep*{weidenschilling-et-al1989}.
Accordingly, there was no single convincing report against the submicrometer-sized grains in dust aggregates that form planetesimals in protoplanetary discs, as far as we know.

What came as a great surprise is that recent studies on the mechanical properties of dust aggregates in protoplanetary discs shed doubt on the consensus about the size of constituent grains (hereafter, monomers) in dust aggregates.
\citet{arakawa-nakamoto2016} claimed the formation of rocky planetesimals through intense cohesion of nanometer-sized silicate monomers that were produced by evaporation of presolar submicrometer-sized silicate grains in protoplanetary discs and subsequent condensation of the vapor.
However, they seem to have overlooked one important evidence that cohesion of submicrometer-sized silicate grains has been one oder of magnitude underestimated in former times \citep{kimura-et-al2015}.
\citet{okamoto-nakamura2017} conducted impact crating experiments on highly porous targets and applied their new empirical scaling law to comet 9P/Tempel 1.
According to their estimates of the tensile strength, the monomers of dust aggregates in 9P/Tempel 1 have large radii of $r_0 = 45$--$1050~\micron$, but it is odd that the monomers comprising dust aggregates are larger than the aggregates with a typical radius of $R \approx 40~\micron$ in 9P/Tempel 1 \citep*[cf.][]{kobayashi-et-al2013}.
On the basis of their numerical simulation on the tensile strength of porous dust aggregates, \citet*{tatsuuma-et-al2019} proposed a radius of $r_0 = 3.3$--$220~\micron$ for the monomers of dust aggregates in 67P/C-G, instead of $r_0 \sim 0.1~\micron$ as commonly believed.
While they provide an empirical formula that reproduces their numerical results, it turned out that if $r_0 = 0.1~\micron$, their formula significantly overestimates the tensile strength estimated for overhangs on the surface of comet 67P/C-G.
In addition, their empirical formula predicts a value that exceeds the tensile strength of dust aggregates consisting of water ice grains with radius $r_0 = 2.38 \pm 1.11\,\micron$ measured in the laboratory by \citet{gundlach-et-al2018}.
To correctly understand the size of grains that form planetesimals in protoplanetary discs, therefore, one must seek a remedy against the apparent disagreement between model predictions and measurements of tensile strengths.

In the field of fracture mechanics, it is well-known that the tensile strength of porous media such as snow, ice, aerogels, minerals, and rocks is weakly dependent on the volume of the media \citep[e.g.][]{sommerfeld1974,petrovic2003,patil-et-al2017,nakamura-et-al2015}.
However, numerical simulations by both \citet*{seizinger-et-al2013} and \citet{tatsuuma-et-al2019} dismissed the idea that the tensile strength of dust aggregates depends on the number of monomers, namely, the volume of dust aggregates.
Here, we cannot help but wonder if the authors of numerical studies failed to notice the volume effect on the tensile strength of dust aggregates, because the volume effect is so subtle that it easily escapes detection.
Therefore, we revisit the tensile strength of dust aggregates to restore the consensus about the size of their constituent grains condensed in protoplanetary discs, by explicitly taking the volume effect into account.

\section{Tensile strength of dust aggregates} \label{sec:tensile-strength}

The so-called JKR theory provides a rigorous solution for the interaction between elastic spheres characterized by elastic properties of a solid, more precisely, the surface energy $\gamma$, Young's modulus $E$, and Poisson's ratio $\nu$ \citep*{johnson-et-al1971}.
\citet{seizinger-et-al2013} and \citet{tatsuuma-et-al2019} performed computer simulations on tensile stress acting on dust aggregates consisting of monodisperse spherical monomers based on the discrete element method (DEM) with the JKR theory.
Since theoretical studies on the tensile strength of dust aggregates have been pursued for more than a century, we shall make full use of an analytical model for the tensile strength of dust aggregates based on the JKR theory of contact mechanics and the Griffith theory of fracture mechanics \citep[e.g.][]{hertz1881,griffith1921,rumpf1970,kendall1987}.
The tensile strength $\sigma$ of a dust aggregate consisting of spherical monomers with the coordination number $n_\mathrm{c}$ may be given by \citep*{kendall1987,kendall-stainton2001,bika-et-al2001}
\begin{eqnarray}
\sigma & = & \phi^\beta n_\mathrm{c} \gamma r_0^{-1} \epsilon^{-1/2} ,
\label{eq:kendall}
\end{eqnarray}
where $\phi$ and $\epsilon$ denote the volume filling factor of the aggregate and the ratio of the maximum flaw size to the diameter of the monomers, respectively.
Note that equation~(\ref{eq:kendall}) reduces to \citeauthor{rumpf1970}'s classical formula if $\beta = 1$ and $\epsilon = \left({4/3}\right)^6$, and to \citeauthor{kendall1987}'s formula if $\beta = 2$ and $n_\mathrm{c} = 17.5\, \phi^2$ \citep{rumpf1970,kendall1987}.
Hereafter we limit our study to dust aggregates of small elastic grains whose compositions are relevant to primitive dust in protoplanetary discs such as water ice, silicates, and organics.

The tensile strength of an aggregate is known to scale with the volume $V$ of the aggregate as $\sigma \propto V^{-1/m}$ where $m$ is commonly referred to as the Weibull modulus \citep[cf.][]{carpinteri1994,petrovic2003}.
Accordingly, we may rewrite equation~(\ref{eq:kendall}) in the following form:
\begin{eqnarray}
\sigma & = & \left({\frac{4\pi N}{3}}\right)^{1/m} \epsilon^{-1/2}  \phi^{\beta-1/m} n_\mathrm{c} \gamma r_0^{3/m-1} V^{-1/m} , 
\label{eq:tensile-strength}
\end{eqnarray}
where $N$ is the number of monomers in the aggregate.
We may regard $\left({4\pi N/3}\right)^{1/m} \epsilon^{-1/2}$ in equation~(\ref{eq:tensile-strength}) to be a constant, because the condition $\epsilon \propto N^{2/m}$ holds for a power-law distribution of flaw sizes \citep{housen-holsapple1999,carpinteri-puzzi2007}.

The Weibull modulus $m$ is known to be material dependent, although $m = 6$ is expected for the fully cracked state, in which the average distance between flaws is equal to the flaw size \citep{housen-holsapple1999,nakamura-et-al2015}.
Hereafter, we shall assume the Weibull modulus $m$ to be $m = 5$ for water ice, $m = 8$ for siliceous material, and $m = 6$ for carbonaceous matter, by taking into account the literature values of $m \approx 5$ for water ice, $m = 6$--$10$ for amorphous silica, and $m = 6.2$ for amorphous diamond-like carbon \citep{petrovic2003,klein2009,borrero-et-al2010}.

To examine whether or not outcomes of computational simulations as well as laboratory experiments are reproduced by equation~(\ref{eq:tensile-strength}), we need a relationship between $n_\mathrm{c}$ and $\phi$.
In general, the coordination number $n_\mathrm{c}$ increases with the volume filling factor $\phi$, although there is no consensus on the formula to describe the relationship \citep[see][for a review]{vanantwerpen-et-al2010}.
For simplicity, we shall use the following relationship between $n_\mathrm{c}$ and $\phi$:
\begin{eqnarray}
n_\mathrm{c} & = & c_1 \exp \left({d_1 \phi}\right) ,
\label{eq:meissner-relation}
\end{eqnarray}
where $c_1 = 2.0$ and $d_1 = 2.4$ were determined by \citet*{meissner-et-al1964}.

\section{Comparison to available data on the tensile strength of dust aggregates}
\label{sec:results}

\subsection{Computer simulations}

\subsubsection{Aggregates of spheres}

\begin{figure}
\centering
\includegraphics[width=0.4\columnwidth]{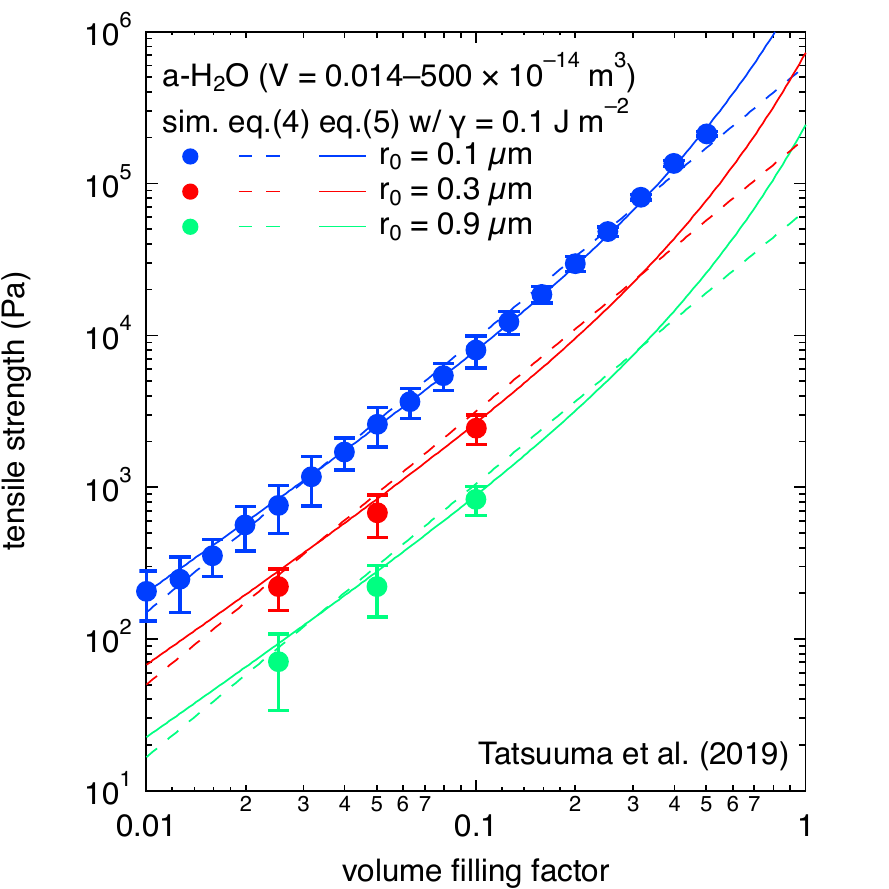}\includegraphics[width=0.4\columnwidth]{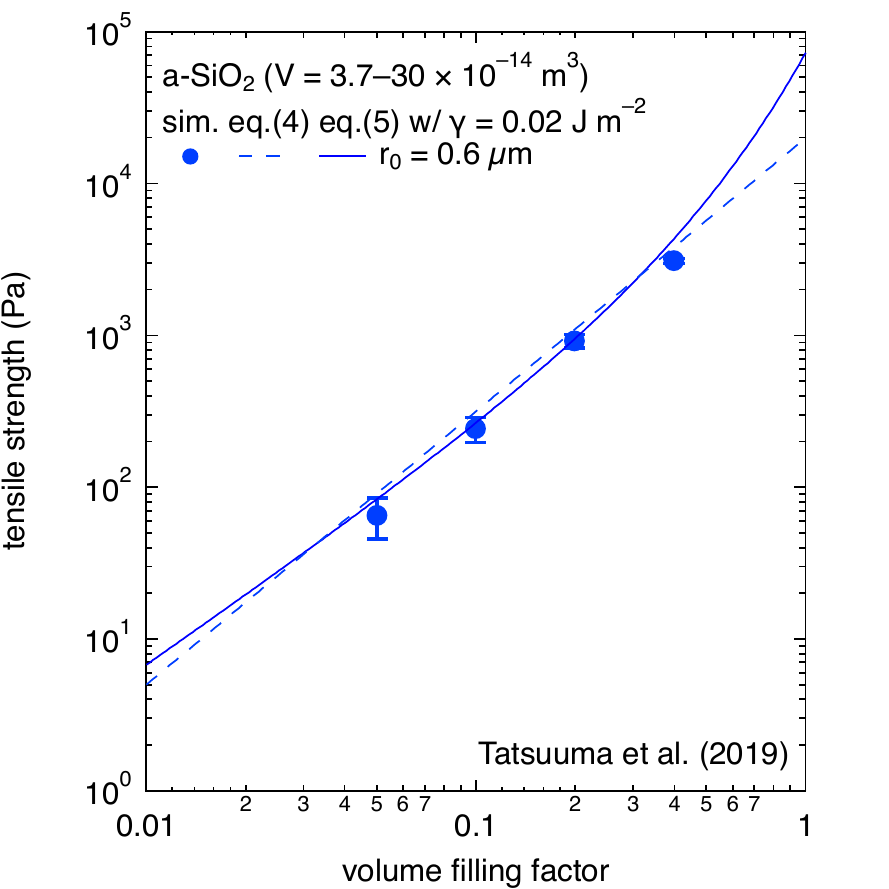}\\
\includegraphics[width=0.4\columnwidth]{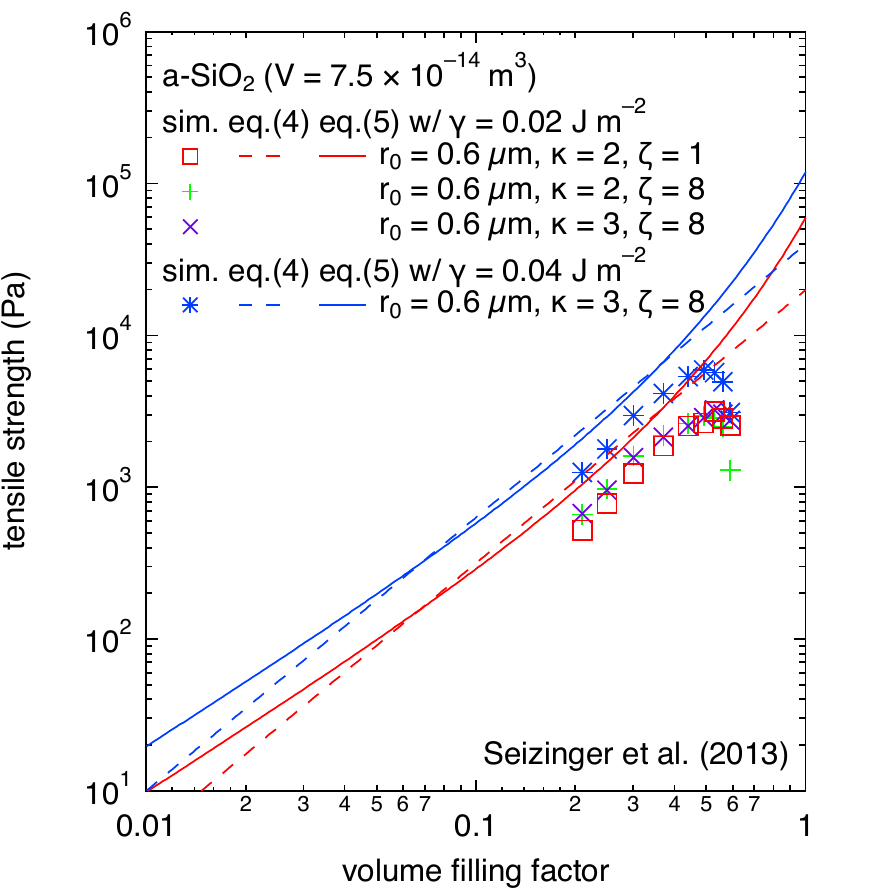}\includegraphics[width=0.4\columnwidth]{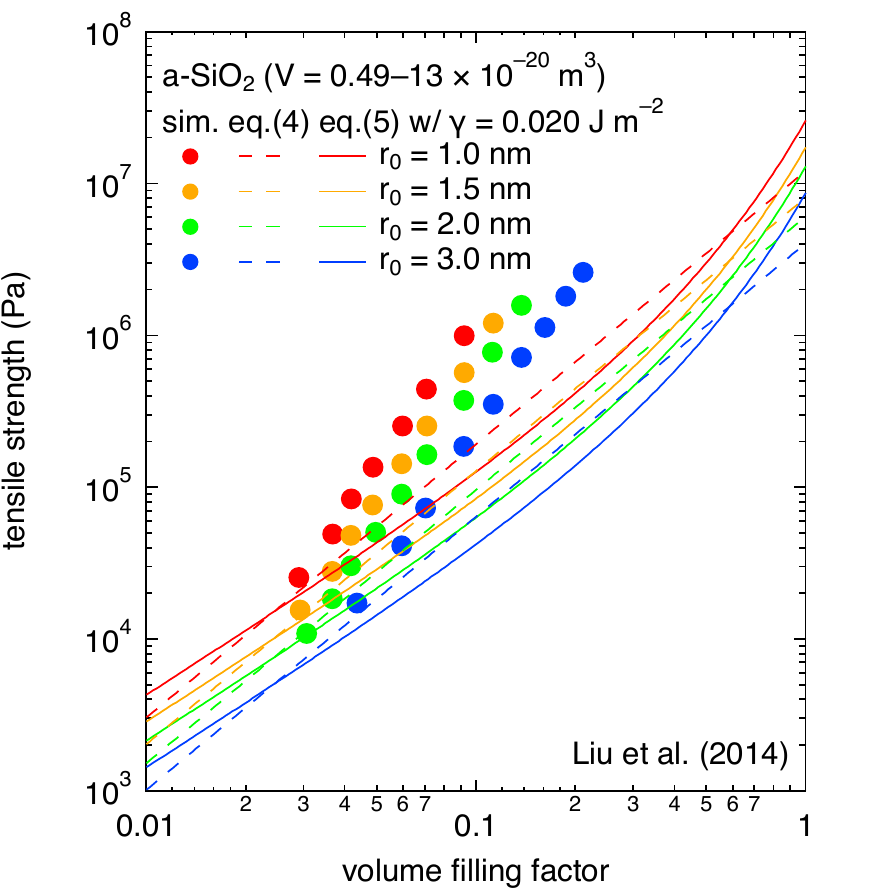}
\caption{A comparison of tensile strength for porous dust aggregates of monodisperse spherical monomers between DEM simulations and our formula. The symbols: computer simulations; the dashed lines: equation~(\ref{eq:tatsuuma}); the solid lines: equation~(\ref{eq:JKR}). Top: simulations based on the JKR theory with $\gamma = 0.1~\mathrm{J~m^{-2}}$ (left) and $\gamma = 0.02~\mathrm{J~m^{-2}}$ (right) by \citet{tatsuuma-et-al2019}; left bottom: simulations based on the JKR theory with $\gamma = 0.02$ and $\gamma = 0.04~\mathrm{J~m^{-2}}$ by \citet{seizinger-et-al2013} where $\kappa$ and $\zeta$ denote their so-called wall-glueing factor and rolling modifier; right bottom: DEM simulations based on a combination of the JKR theory and the DMT theory by \citet{liu-et-al2014}. 
\label{fig:DEMsimulations}}
\end{figure}

On the basis of their DEM simulations, \citet{tatsuuma-et-al2019} proposed that the tensile strength $\sigma$ of porous dust aggregates consisting of $N$ identical spherical monomers is given by the following empirical formula:
\begin{eqnarray}
\sigma & = & \sigma_0 \left({\frac{\gamma}{0.1~\mathrm{J~m^{-2}}}}\right) \left({\frac{r_0}{0.1~\micron} }\right)^{-1} \left({\frac{\phi}{0.1}}\right)^{\beta} ,
\label{eq:tatsuuma}
\end{eqnarray}
with $\sigma_0 = 9.51~\mathrm{kPa}$ and $\beta = 1.8$, regardless of $N$.
By the same token, an empirical formula for the tensile strength of porous dust aggregates found by \citet{seizinger-et-al2013} corresponds to $\sigma_0 = 4.43~\mathrm{kPa}$ and $\beta = 1.88$ in their DEM simulations.
One might attribute the difference in the $\sigma_0$ and $\beta$ values of the empirical formula between these two groups to the uncertainty of the results by DEM simulations.
Here, we suppose that the results of \citet{tatsuuma-et-al2019} are more accurate, compared with those of \citet{seizinger-et-al2013}, according to the magnitude of smaller time steps used in the former than the latter.
\citet{tatsuuma-et-al2019} fixed the number of monomers to $N = 2^{14}$ and the volume of the aggregate varies with the volume filling factor $\phi$, while \citet{seizinger-et-al2013} fixed the volume to $V = 7.5 \times{10}^{-14}~\mathrm{m^3}$ and the number of monomers varies with $\phi$.
If we consider an aggregate with $r_0 = 0.6~\micron$ and $\phi = 0.19765$, then we have $N = 2^{14}$ and $V = 7.5 \times{10}^{-14}~\mathrm{m^3}$ in both \citet{tatsuuma-et-al2019} and \citet{seizinger-et-al2013}.
By inserting $r_0 = 0.6~\micron$, $\phi = 0.19765$, and $\gamma = 0.02~\mathrm{J~m^{-2}}$ into equation~(\ref{eq:tatsuuma}), we obtain $\sigma = 1.14~\mathrm{kPa}$ for the former ($\sigma_0 = 10~\mathrm{kPa}$, $\beta = 1.8$) and $\sigma = 0.532~\mathrm{kPa}$ for the latter ($\sigma_0 = 4.43~\mathrm{kPa}$, $\beta = 1.88$).
Accordingly, we may consider that the tensile strengths of dust aggregates determined in the DEM simulations by \citet{seizinger-et-al2013} are underestimated by a factor of 2.

What follows is the best fit of equation~(\ref{eq:tensile-strength}) to numerical results of tensile strengths by \citet{tatsuuma-et-al2019} with $\gamma = 0.1~\mathrm{J~m^{-2}}$, $r_0 = 0.1~\micron$, and $N = 2^{14}$:
\begin{eqnarray}
\sigma & = & 8~\mathrm{kPa}\, \left({\frac{\gamma}{0.1~\mathrm{J~m^{-2}}}}\right) \left({\frac{r_0}{0.1~\micron} }\right)^{3/m-1} \left({\frac{\phi}{0.1}}\right)^{\beta-1/m} \exp\left[{\alpha\left({\frac{\phi}{0.1}-1}\right)}\right] \left({\frac{V}{686~\micron^3}}\right)^{-1/m} ,
\label{eq:JKR}
\end{eqnarray}
with $\beta = 1.5$ and $\alpha = 0.24$, implying $\epsilon = 10^2$ for $N = 2^{14}$.
It should be noted that equation~(\ref{eq:JKR}) cannot be in principle applied to highly compact aggregates of $\phi \ga 0.74$, because the maximum value of volume filling factor for aggregates of spherical monomers is $\phi = \sqrt{2}\pi/6$ \citep{kepler1611}.
The top panels of Fig.~\ref{fig:DEMsimulations} demonstrate that equation~(\ref{eq:JKR}; the solid lines) fairly well reproduces numerical results of \citet[][the filled circles]{tatsuuma-et-al2019} as well as their empirical formula of equation~(\ref{eq:tatsuuma}; the dashed lines), irrespective of the values assumed for monomer radius $r_0$ and surface energy $\gamma$.
These results validate equation~(\ref{eq:JKR}) as a substitute for equation~(\ref{eq:tatsuuma}), which has been formulated by \citet{tatsuuma-et-al2019} for their numerical simulations.

\citet{seizinger-et-al2013} was ahead of \citet{tatsuuma-et-al2019} concerning DEM-based numerical studies on the tensile strength of porous dust aggregates consisting of monodisperse spherical monomers.
DEM simulations performed by \citet{seizinger-et-al2013} are based on the JKR theory, although adhesion of monomers to two plates, in which the aggregates are sandwiched, was artificially increased by a factor $\kappa$.
They investigated how the tensile strength of the aggregates varies with the volume filling factor and the volume of the aggregates as well as the radius of monomers. 
The left bottom panel of Fig.~\ref{fig:DEMsimulations} shows that equation~(\ref{eq:JKR}) predicts the tensile strength of dust aggregates that exceeds the numerical results of \citet{seizinger-et-al2013} by a factor of 2, similar to equation~(\ref{eq:tatsuuma}), but provides slightly better fits to the results than equation~(\ref{eq:tatsuuma}) does.

DEM-based numerical studies by \citet{liu-et-al2014} utilized a generalized model for the contact of elastic solids proposed by \citet{schwarz2003}, which incorporates a short-range force in the JKR theory and a long-range force in the DMT theory.
We assume the predominance of long-range forces on tensile stress and thus take $3/4$ times the long-range component of surface energy to interpret their results in the framework of the JKR theory.
As plotted in the right bottom panel of Fig.~\ref{fig:DEMsimulations}, the values of tensile strength in their results at the smallest volume filling factors appear to be on the same order of magnitude as the values predicted by equation~(\ref{eq:JKR}), while the deviations grows with the volume filling factor of the aggregates.
The dependences of tensile strength on the volume filling factor and the monomer's radius are also stronger in the DEM simulations by \citet{liu-et-al2014} than those by \citet{tatsuuma-et-al2019} and \citet{seizinger-et-al2013} (cf. the top and left-bottom panels of Fig.~\ref{fig:DEMsimulations}).
The strong dependence of tensile strength on the volume filling factor in \citet{liu-et-al2014} most likely originates from an increasing contribution of short-range forces for compact aggregates, because the contribution of short-range forces to tensile stress should increase with the volume filling factor.

\subsection{Laboratory experiments}

\subsubsection{Water ice}

\begin{figure}
\centering\includegraphics[width=0.4\columnwidth]{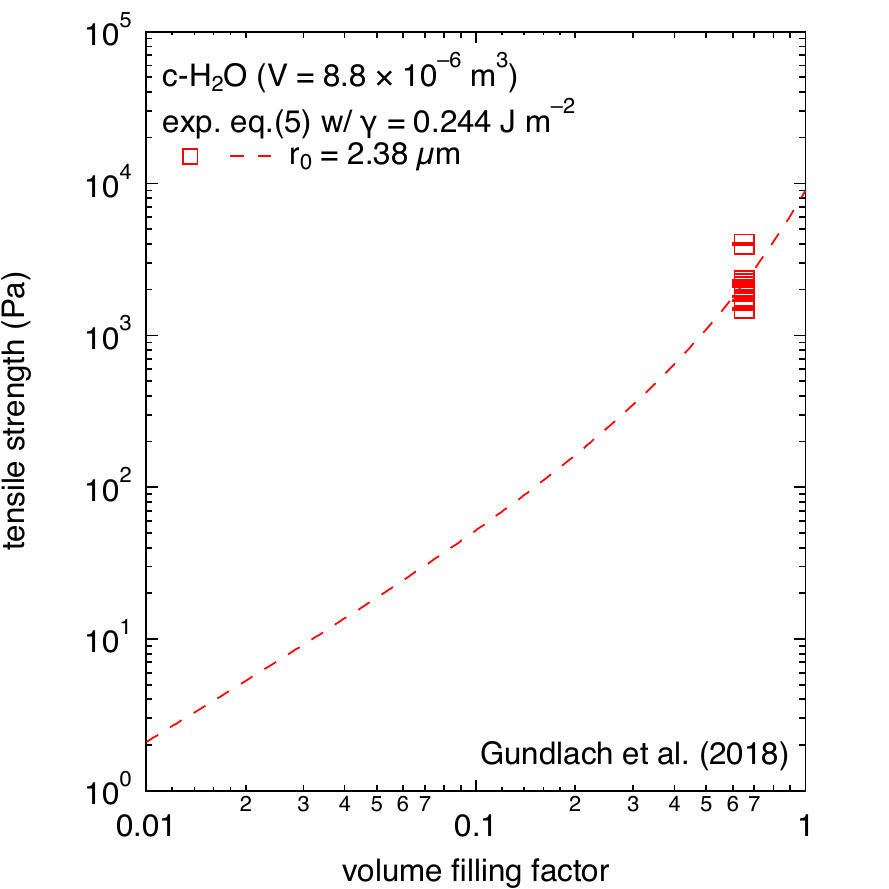}
\caption{A comparison of tensile strength for dust aggregates of monodisperse spherical ice monomers between laboratory experiments and our formula. The open squares: experimental data for crystalline water ice in air by \citet{gundlach-et-al2018}; the dashed line: equation~(\ref{eq:JKR}) with $\gamma = 0.244~\mathrm{J~m^{-2}}$ for crystalline water ice.
\label{fig:experiments-ice}}
\end{figure}

\citet{gundlach-et-al2018} measured the tensile strength of compact dust aggregates consisting of polydisperse spherical water ice particles with $r_0 = 2.38 \pm 1.11~\micron$ at a temperature of $150~\mathrm{K}$ in air.
They produced crystalline water ice particles by spraying water droplets into liquid nitrogen and formed porous dust aggregates by pressing the monomers into a cylinder.
We shall first compare equation~(\ref{eq:JKR}) to the tensile strengths measured by \citet{gundlach-et-al2018}, which were found to be much lower than equation~(\ref{eq:tatsuuma}) by \citet{tatsuuma-et-al2019}.
Throughout the paper, we assume $\gamma = 0.244~\mathrm{J~m^{-2}}$ for the surface energy of crystalline water ice Ih derived from the density-functional theory (DFT), although the value may differ by 20\%, depending on the crystal face \citep{pan-et-al2010}.
Figure~\ref{fig:experiments-ice} shows that equation~(\ref{eq:JKR}) agrees with laboratory experiments on the tensile strengths of dust aggregates composed of crystalline water ice, dissimilar to equation~(\ref{eq:tatsuuma}) proposed by \citet{tatsuuma-et-al2019}.
This demonstrates that the volume effect on the tensile strength incorporated in equation~(\ref{eq:JKR}) provides a remedy for the discrepancy between laboratory experiments and model predictions.

\subsubsection{Silicates}

\begin{figure}
\centering
\includegraphics[width=0.4\columnwidth]{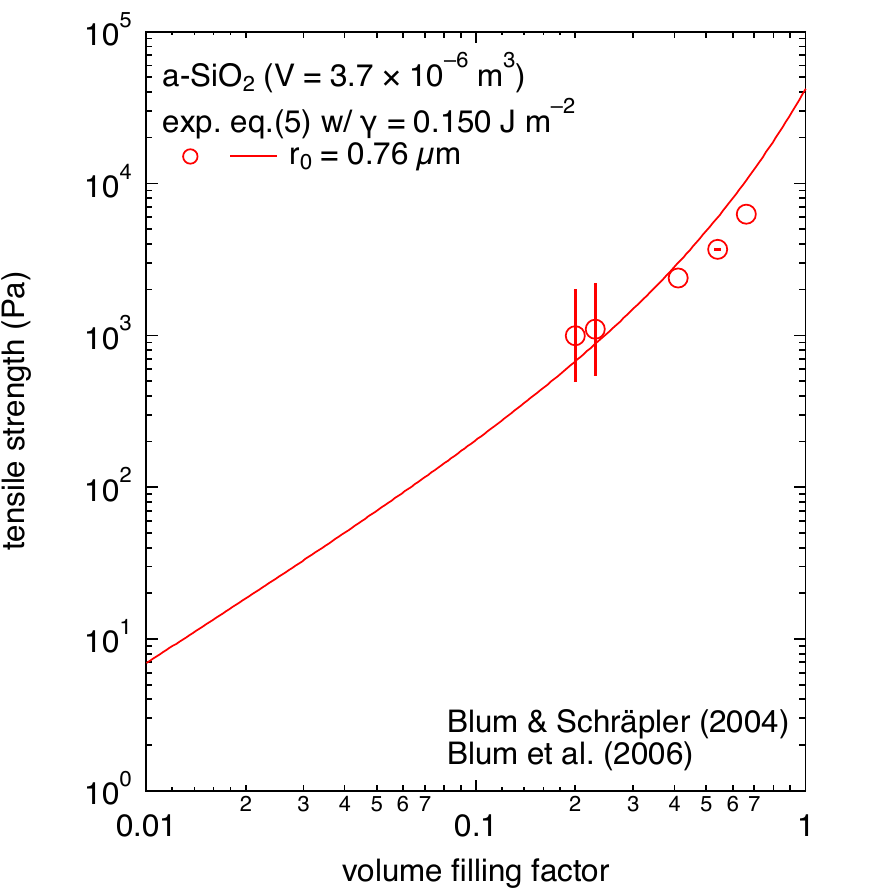}\includegraphics[width=0.4\columnwidth]{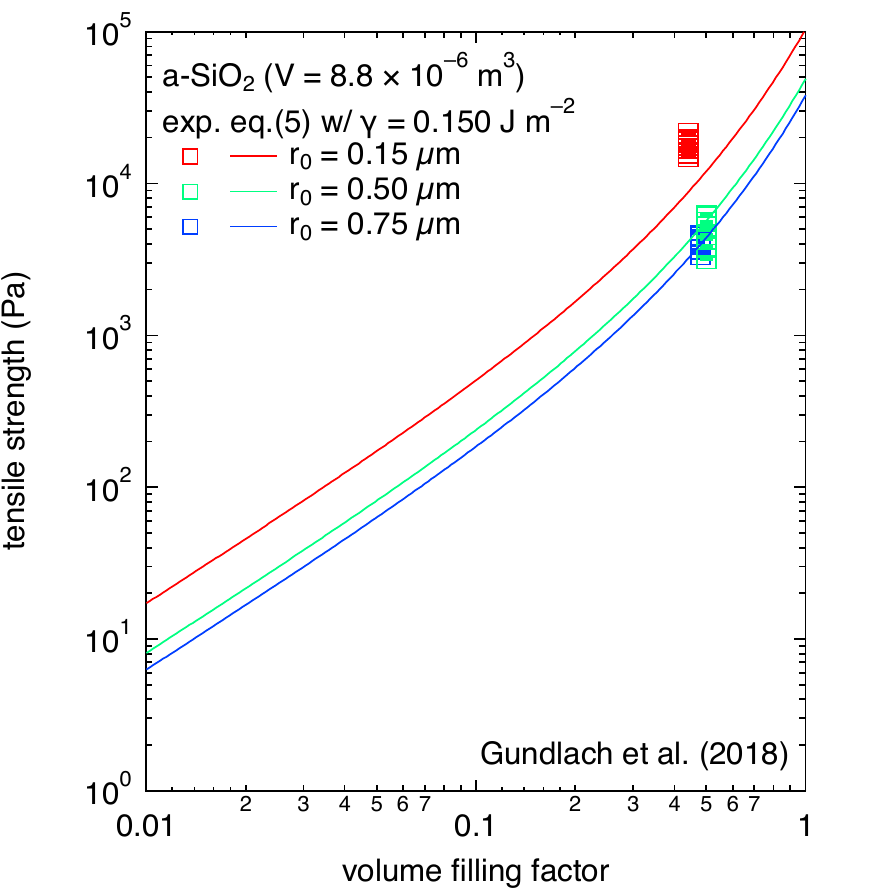}\\
\includegraphics[width=0.4\columnwidth]{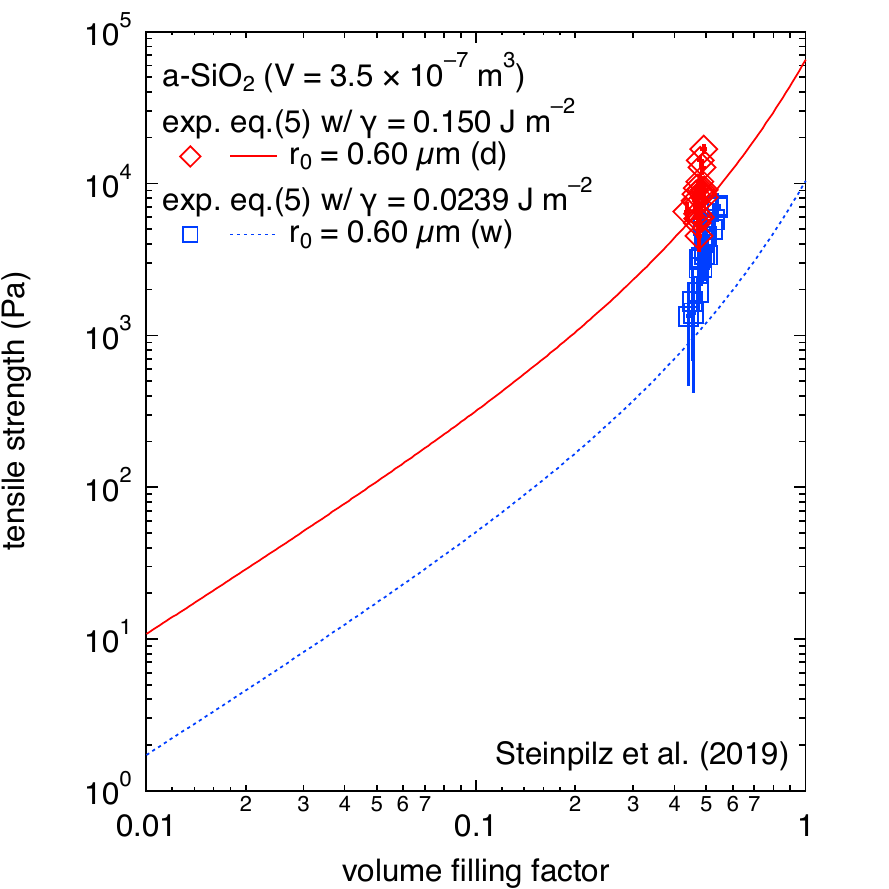}\includegraphics[width=0.4\columnwidth]{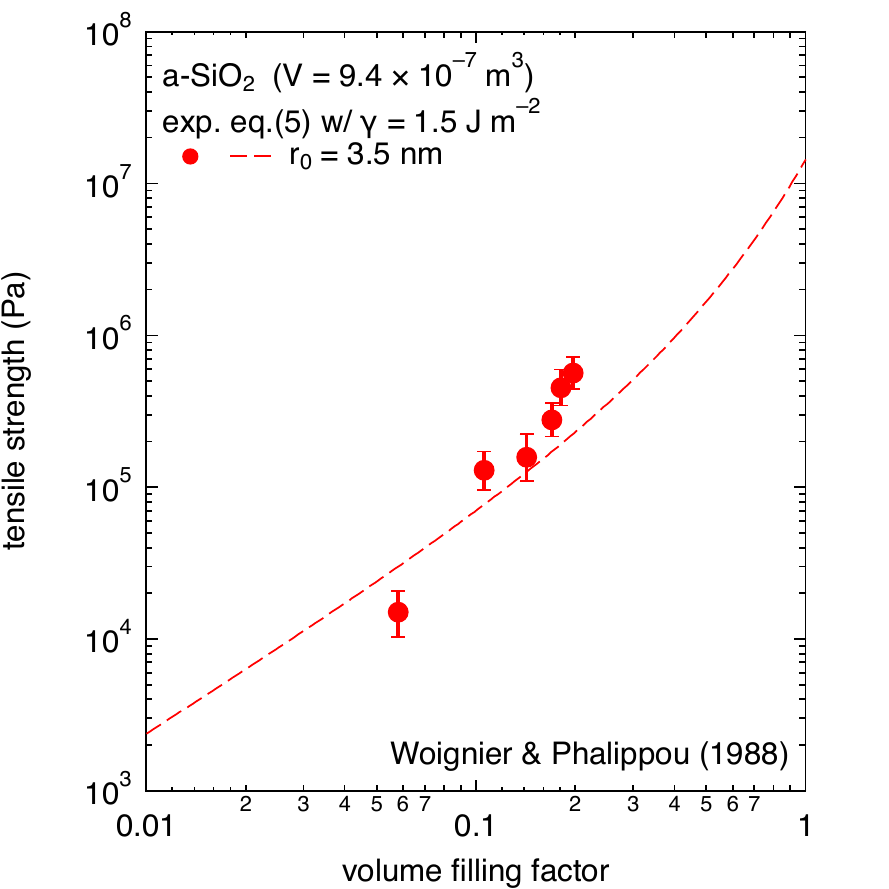}
\caption{A comparison of tensile strength for dust aggregates of monodisperse and polydisperse spherical silica monomers between laboratory experiments and our formula. The solid lines: equation~(\ref{eq:JKR}) with $\gamma = 0.150~\mathrm{J~m^{-2}}$ for $\mathrm{sicastar}\textsuperscript{\textregistered}$ silanol-bonded silica in vacuum or high temperature; the dashed line: equation~(\ref{eq:JKR}) with $\gamma = 1.5~\mathrm{J~m^{-2}}$ for siloxane-bonded amorphous silica; the dotted line: equation~(\ref{eq:JKR}) with $\gamma = 0.0239~\mathrm{J~m^{-2}}$ for hydrated amorphous silica in air. The open circles (left top): experimental data for porous and compact aggregates of amorphous silica measured in vacuum by \citet{blum-schraepler2004} and \citet{blum-et-al2006}; the open squares (right top): experimental data for amorphous silica in air by \citet{gundlach-et-al2018}; the open squares (left bottom): experimental data for compact aggregates of wet (w), unheated amorphous silica measured in air by \citet*{steinpilz-et-al2019}; the open diamonds (left bottom): experimental data for compact aggregates of dry (d), heated amorphous silica measured in air by \citet{steinpilz-et-al2019}; the filled circles (right bottom) experimental data for amorphous silica in air by \citet{woignier-phalippou1988}.
\label{fig:experiments-sphere}}
\end{figure}

\citet{blum-schraepler2004} and \citet{blum-et-al2006} used amorphous silica ``$\mathrm{sicastar}\textsuperscript{\textregistered}$'' particles with radius $r_0 = 0.76~\micron$ produced by micromod Partikeltechnologie GmbH to form porous ($\phi \approx 0.15$--$0.33$) dust aggregates.
They uni-axially compressed the aggregates of monodisperse spheres to a pressure of $(4 \pm 2) \times {10}^{3}~\mathrm{Pa}$ prior to their measurements of tensile strength at medium vacuum conditions ($\sim 100~\mathrm{Pa}$).
\citet{blum-et-al2006} also produced compact ($\phi = 0.41$--$0.66$) dust aggregates of monodisperse $\mathrm{sicastar}\textsuperscript{\textregistered}$ spheres by applying an omnidirectional pressure to the aggregates.
We assume $\gamma = 0.150~\mathrm{J~m^{-2}}$ for the surface energy of $\mathrm{sicastar}\textsuperscript{\textregistered}$ (micromod Partikeltechnologie GmbH), which is consistent with collision experiments using $\mathrm{sicastar}\textsuperscript{\textregistered}$ spheres \citep[cf.][]{kimura-et-al2015}.
The left top panel of Fig.~\ref{fig:experiments-sphere} shows that the experimental data for the tensile strength of porous and compact aggregates obtained by \citet{blum-schraepler2004} and \citet{blum-et-al2006} are reasonably in good harmony with equation~(\ref{eq:JKR}) if $\gamma = 0.150~\mathrm{J~m^{-2}}$. 
It should be noted that equation~(\ref{eq:JKR}) shows to some extent deviations from experimental data on the tensile strength of compact dust aggregates with large $\phi$ values.

\citet{gundlach-et-al2018} measured the tensile strength of compact dust aggregates consisting of monodisperse $\mathrm{sicastar}\textsuperscript{\textregistered}$ spheres at room temperature in air.
Since they formed compact dust aggregates by pressing the monomers into a cylinder, we consider that monomers are in contact without help of adsorbed water molecules on their surfaces (i.e. $\gamma = 0.150~\mathrm{J~m^{-2}}$).
The right top of Fig.~\ref{fig:experiments-sphere} compares equation~(\ref{eq:JKR}) with their laboratory experiments on the tensile strengths of dust aggregates composed of amorphous silica.
The tensile strength of compact dust aggregates consisting of amorphous silica monomers with $r_0 = 0.15~\micron$ measured in air by \citet{gundlach-et-al2018} is higher than equation~(\ref{eq:JKR}) with $\gamma = 0.150~\mathrm{J~m^{-2}}$, although their results with $r_0 = 0.50$ and $0.75~\micron$ are in good agreement with equation~(\ref{eq:JKR}).

\citet{steinpilz-et-al2019} also used $\mathrm{sicastar}\textsuperscript{\textregistered}$ to form compact dust aggregates of silica spheres with $r_0 = 0.6~\micron$ by pressing the aggregates up to a pressure of $5.5 \times {10}^{4}~\mathrm{Pa}$\footnote{\citet{steinpilz-et-al2019} did not explicitly describe the quantity of applied pressures to press their aggregates, but mentioned that they took the same procedure as \citet*{meisner-et-al2012} who gave a pressure of $\la 5.5 \times {10}^{4}~\mathrm{Pa}$.}.
They measured the tensile strength of compact aggregates at room temperature in air before and after heating to $250^\circ\mathrm{C}$ for $24~\mathrm{hrs}$ in the oven.
Because of its hydrophilic nature, amorphous silica particles in air are known to swell with adsorbed water molecules and the surface energy of amorphous silica is reduced typically to $\gamma \sim 0.0239~\mathrm{J~m^{-2}}$ at room temperature \citep*{kendall-et-al1987}.
Therefore, evaporation of water molecules by heating to higher temperatures elevates the surface energy of amorphous silica, in a similar way to vacuum conditions \citep{maszara-et-al1988,kimura-et-al2015}.
By the same token, \citet{kamiya-et-al2002} observed an increase in the tensile strength of amorphous silica powders with temperature, as expected from equation~(\ref{eq:JKR}), which predicts the proportionality of tensile strength to surface energy.
The left bottom panel of Fig.~\ref{fig:experiments-sphere} shows that the tensile strengths of unheated and heated compact aggregates are consistent with equation~(\ref{eq:JKR}) if $\gamma = 0.0239~\mathrm{J~m^{-2}}$ for the former and $\gamma = 0.150~\mathrm{J~m^{-2}}$ for the latter.
Note that \citet{steinpilz-et-al2019} corrected the volume filling factor for unheated aggregates by taking into account the apparent reduction in the volume filling factor due to adsorption of water molecules (cf. Appendix~\ref{appendix:water}).
This indicates that lower values of volume filling factor correspond to higher amounts of adsorbed water molecules, which are equivalent to lower values of surface energy.
Accordingly, their results are consistent with equation~(\ref{eq:JKR}), because a reduction in the volume filling factor, in turn, the surface energy, is expected to decrease the tensile strength.

The tensile strength of silica aerogels consisting of polydisperse spherical monomers with radius $r_0 = 3.0$--$4.0~\mathrm{nm}$ was measured at room temperature in air by \citet{woignier-phalippou1988}.
Silica aerogels are highly porous dust aggregates and the monomers are strongly bonded by siloxane ($\ce{Si-O-Si}$) bridges, although silanol ($\ce{Si-OH}$) groupes may remain on the outside.
Therefore, we may apply the value of $\gamma = 1.5~\mathrm{J~m^{-2}}$ to the surface energy of silica aerogels in equation~(\ref{eq:JKR}), irrespective of vacuum conditions, while the value of $\gamma$ for amorphous silica with siloxane bonding is uncertain within a factor of 2 \citep[see][]{kimura-et-al2015}.
The right bottom panel of Fig.~\ref{fig:experiments-sphere} proves that the tensile strength of dust aggregates given by equation~(\ref{eq:JKR}) is applicable to highly porous silica aerogels within a factor of 2, although the dependence of tensile strength on the volume filling factor of dust aggregates appears to be slightly steeper in experiments, compared with equation~(\ref{eq:JKR}).

\begin{figure}
\centering
\includegraphics[width=0.4\columnwidth]{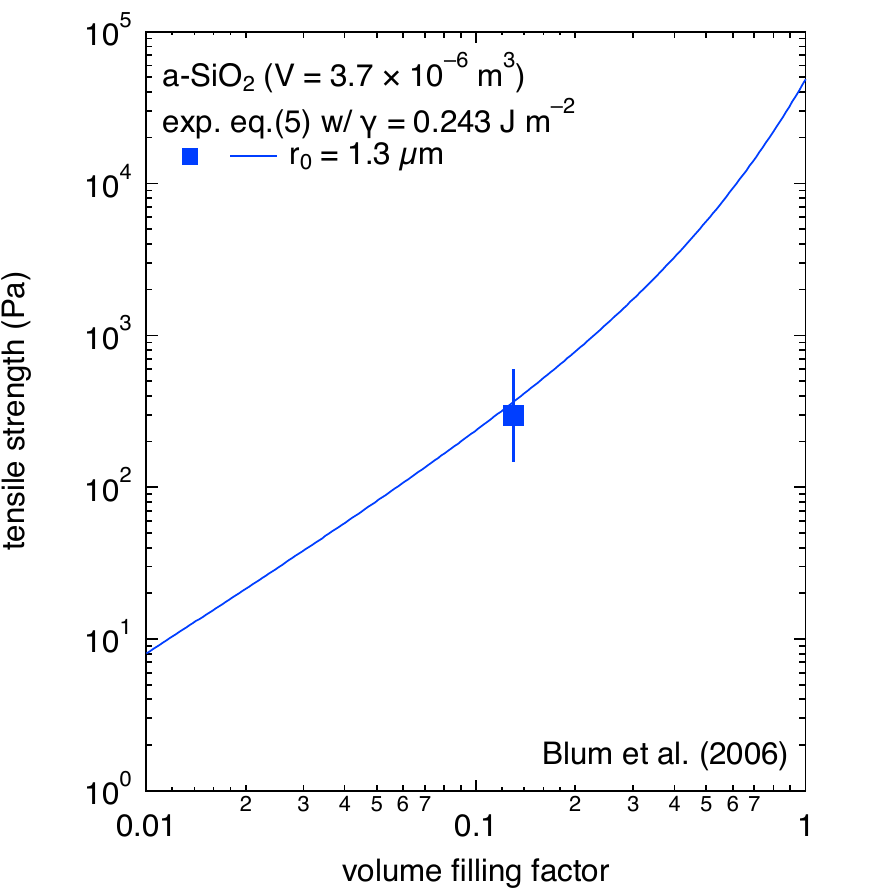}\includegraphics[width=0.4\columnwidth]{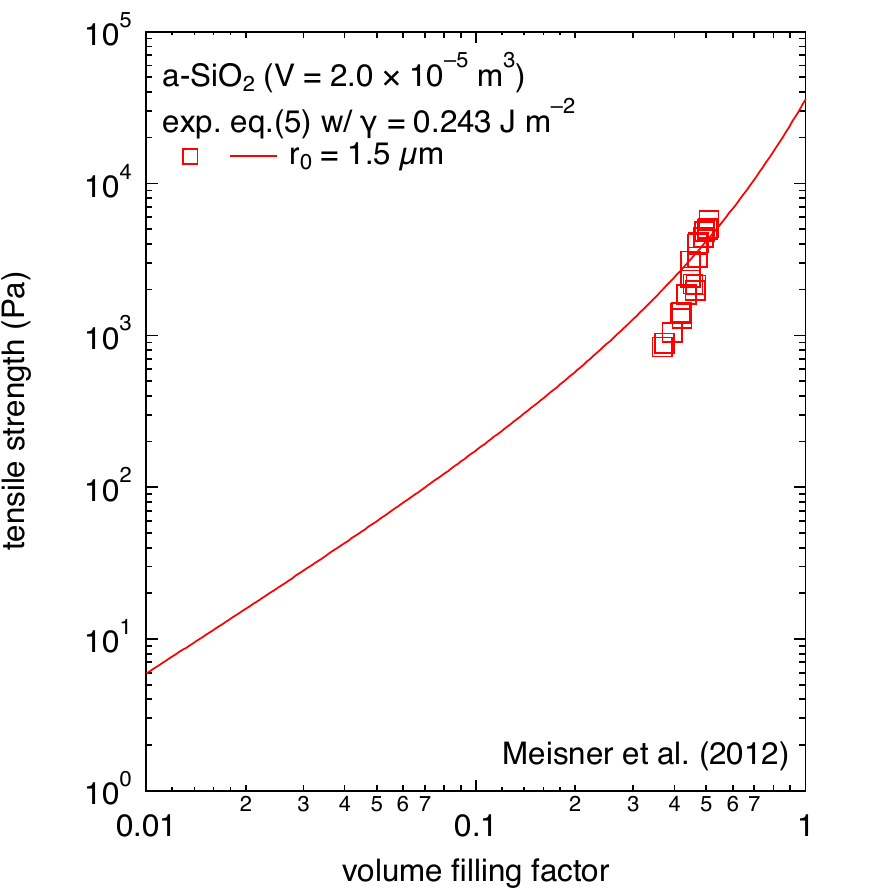}\\
\includegraphics[width=0.4\columnwidth]{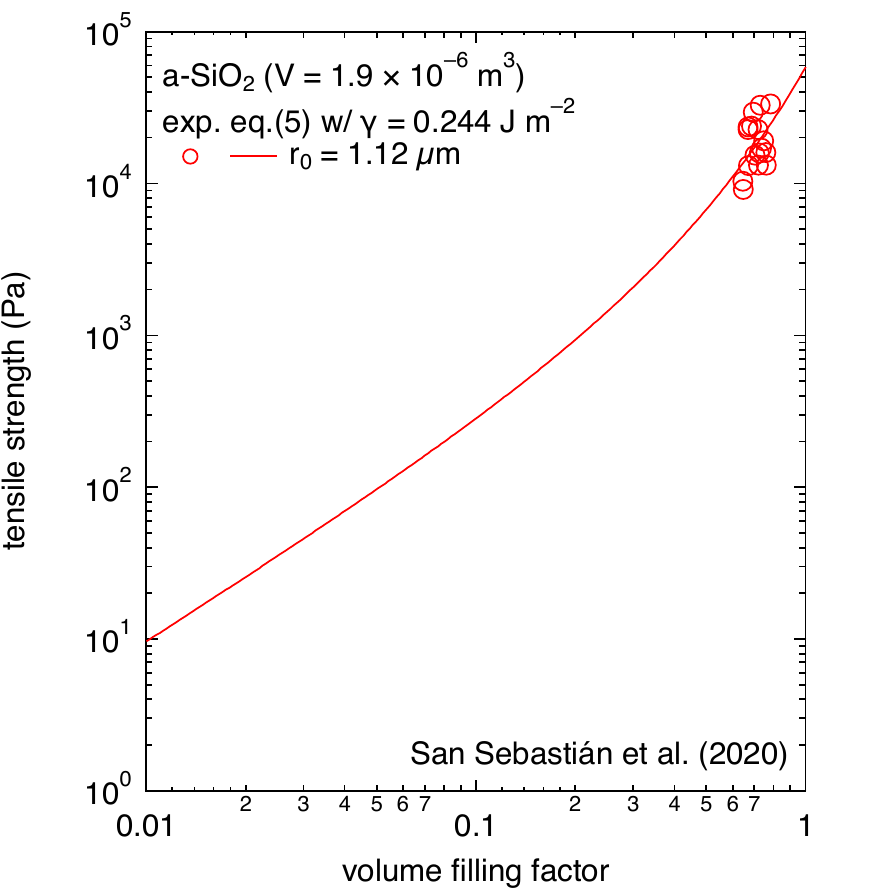}\\
\includegraphics[width=0.4\columnwidth]{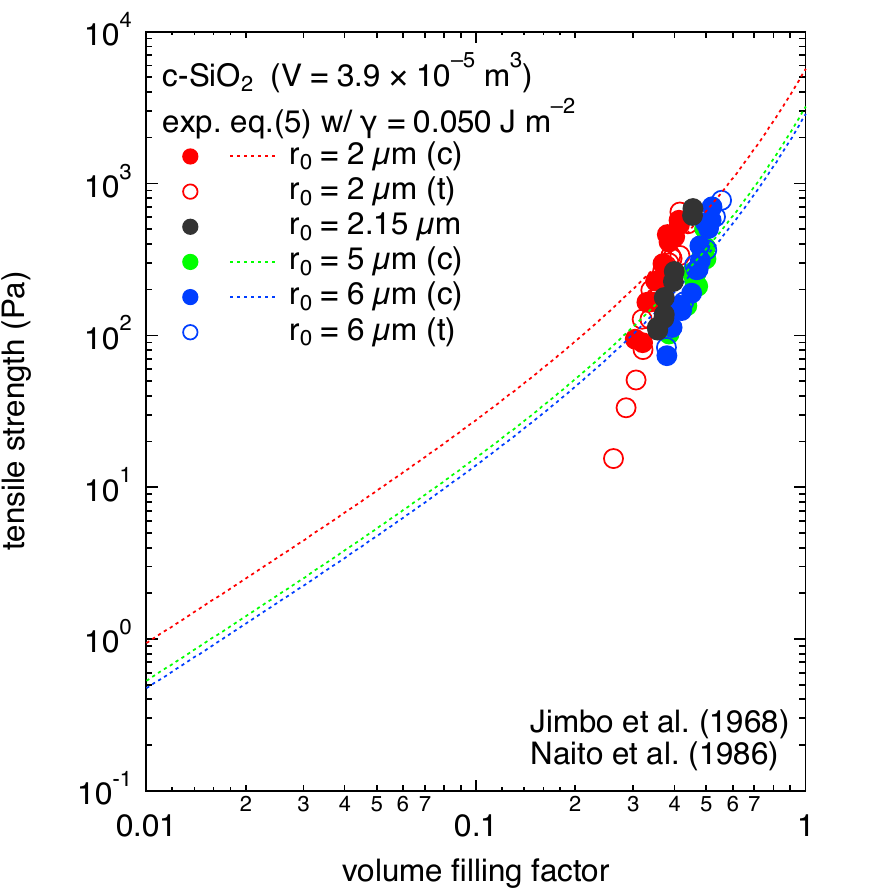}\includegraphics[width=0.4\columnwidth]{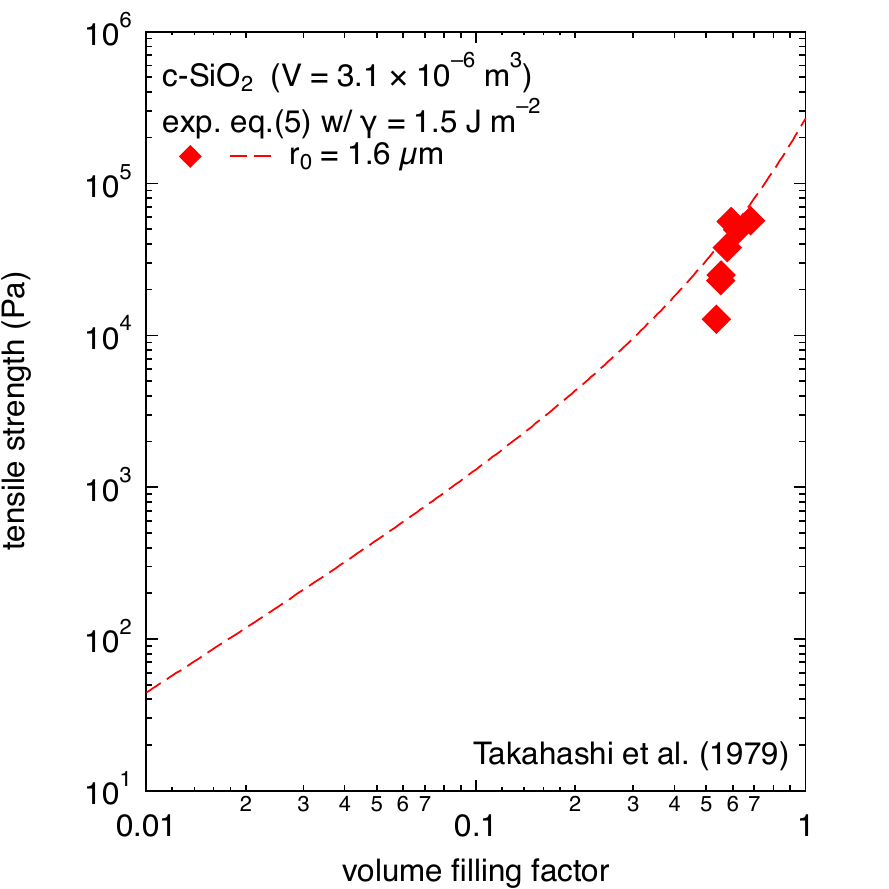}
\caption{A comparison of tensile strength for porous and compact dust aggregates of polydisperse irregularly shaped silica monomers between laboratory experiments and our formula. Left top: porous aggregates of amorphous silica in vacuum by \citet[][the filled squares]{blum-et-al2006}; right top: mildly compact aggregates of amorphous silica in air by \citet[][the open squares]{meisner-et-al2012}; middle: highly compact aggregates of amorphous silica in air by \citet[][the open circles]{sansebastian-et-al2020}; left bottom: silica sand powder beds prepared by compression (c) or tapping (t) in air by \citet*{jimbo-et-al1968} and \citet[][the open and filled circles]{naito-et-al1986}; right bottom: silica sand powder beds prepared by heating and strong compression in air by \citet[][the filled diamonds]{takahashi-et-al1979}; the solid lines: equation~(\ref{eq:JKR}) with $\gamma = 0.243~\mathrm{J~m^{-2}}$ for amorphous silica and equation~(\ref{eq:JKR}) with $\gamma = 0.050~\mathrm{J~m^{-2}}$ or $\gamma = 1.5~\mathrm{J~m^{-2}}$ for crystalline silica.
\label{fig:experiments-irregular}}
\end{figure}

\citet{blum-et-al2006} measured the tensile strengths of porous ($\phi = 0.13$) dust aggregates consisting of polydisperse irregularly shaped silica particles with $r_0 = 1.25^{+3.75}_{-1.20}~\micron$ at medium vacuum conditions ($\sim 100~\mathrm{Pa}$). 
\citet{meisner-et-al2012} prepared more compact ($\phi = 0.37$--$0.51$) dust aggregates of polydisperse irregularly shaped silica particles with  $r_0 = (1.5 \pm 1.0)~\micron$ by applying an omnidirectional pressure of $\la 5.5 \times {10}^{4}~\mathrm{Pa}$ to the aggregates.
They applied the Brazilian test, which is one of the most popular indirect tensile tests, to measure the tensile strength of the aggregates in a vacuum chamber at medium vacuum conditions of $\la 4~\mathrm{Pa}$.
We consider $\gamma = 0.243~\mathrm{J~m^{-2}}$ to best represent the surface energy of amorphous silica in vacuum, since the surface tension of silica glass asymptotically approaches this value at absolute zero \citep{kimura-et-al2020}.
As shown in the top panels of Fig.~\ref{fig:experiments-irregular}, equation~(\ref{eq:JKR}) with $\gamma = 0.243~\mathrm{J~m^{-2}}$ reproduces the tensile strengths of both porous (left) and compact (right) aggregates consisting of irregularly shaped silica particles.
The Brazilian test with highly compact ($\phi = 0.64$--$0.78$) dust aggregates of polydisperse irregularly shaped silica particles was conducted by \citet{sansebastian-et-al2020}.
While the tensile strength of the aggregates was measured at atmospheric conditions, the influence of adsorbed water molecules on the surface of silica particles is most likely negligible for such compact aggregates formed by intense compression prior to the measurements.
Therefore, we compare their experimental results to equation~(\ref{eq:JKR}) with $\gamma = 0.243~\mathrm{J~m^{-2}}$, which is shown in the middle panel of Fig.~\ref{fig:experiments-irregular}.
The good fit of the experimental data to the theoretical curve justifies the validity of equation~(\ref{eq:JKR}) within the expected range of volume filling factor for dust aggregates, namely, $\phi < 0.8$.

\citet{jimbo-et-al1968} and \citet{naito-et-al1986} measured the tensile strength of granular materials consisting of polydisperse irregularly shaped silica sand with radii $r_0 = 2$, $2.15$, $5$, and $6~\micron$.
\citet{jimbo-et-al1968} prepared their silica powder beds by either tapping or compression in the range of $5.0 \times {10}^{2}$--$5.0 \times {10}^{4}~\mathrm{Pa}$, but the tensile strength did not strongly depend on the method of sample preparation as shown in Fig.~\ref{fig:experiments-irregular}.
They determined adhesion forces of the same silica sand powders to flat plates using centrifugal forces, which results in $\gamma = 0.050~\mathrm{J~m^{-2}}$ \citep{asakawa-jimbo1967}\footnote{\citet{asakawa-jimbo1967} derived an average adhesive force of $F_\mathrm{ad} = 3.9 \times {10}^{-6}~\mathrm{N}$ from their measurements with silica sand powders of the mean radius $r_0 = 8.25~\micron$. By applying the JKR theory (i.e. $F_\mathrm{ad} = 3 \pi \gamma r_0$) to their measurements, one could obtain the surface energy of $\gamma = 0.050~\mathrm{J~m^{-2}}$.}.
Therefore, we consider that the value of $\gamma = 0.050~\mathrm{J~m^{-2}}$ is appropriate to the surface energy of quartz used in their experiments, while $\gamma = 1.5~\mathrm{J~m^{-2}}$ for quartz in vacuum (see Appendix~\ref{appendix:quartz} for the surface energy of quartz).
Indeed, their results on the tensile strength of silica sand powder beds are consistent with equation~(\ref{eq:JKR}) if $\gamma = 0.050~\mathrm{J~m^{-2}}$, as shown in the left bottom panel of Fig.~\ref{fig:experiments-irregular}.

\citet{takahashi-et-al1979} heated silica sand of $r_0 = 1.6~\micron$ at a temperature of $110$--$150^\circ\mathrm{C}$ for $48~\mathrm{hrs}$ in air and kept the powders in a desiccator for more than a day prior to their experiments.
It is well-known that the surface energy of crystalline silica (i.e. quartz) increases with temperature, owing to evaporation of water molecules and the formation of siloxane bonding \citep[e.g.][]{parks1984}.
By applying high pressures of $9.8 \times {10}^{4}$--$9.8 \times {10}^{7}~\mathrm{Pa}$ to their fine powder beds, they prepared compact powder beds of irregularly shaped quartz particles. 
They found a gradual increase of tensile strength with pressure up to $4.9 \times {10}^{7}~\mathrm{Pa}$, but no more increase above $4.9 \times {10}^{7}~\mathrm{Pa}$.
The right bottom panel of Fig.~\ref{fig:experiments-irregular} shows that their results are in harmony with equation~(\ref{eq:JKR}) if $\gamma = 1.5~\mathrm{J~m^{-2}}$.
It should be noted that the surface energy of $\gamma \approx 1.5~\mathrm{J~m^{-2}}$ suggests siloxane bonding, which corresponds to the surface chemistry of quartz in vacuum.
Therefore, we attribute their results of high tensile strengths for highly compressed, heated powder beds to the achievement of siloxane bonding between monomers \citep*[see, e.g.][]{stengl-et-al1989}.

\subsubsection{Organic matter}

\begin{figure}
\centering
\includegraphics[width=0.4\columnwidth]{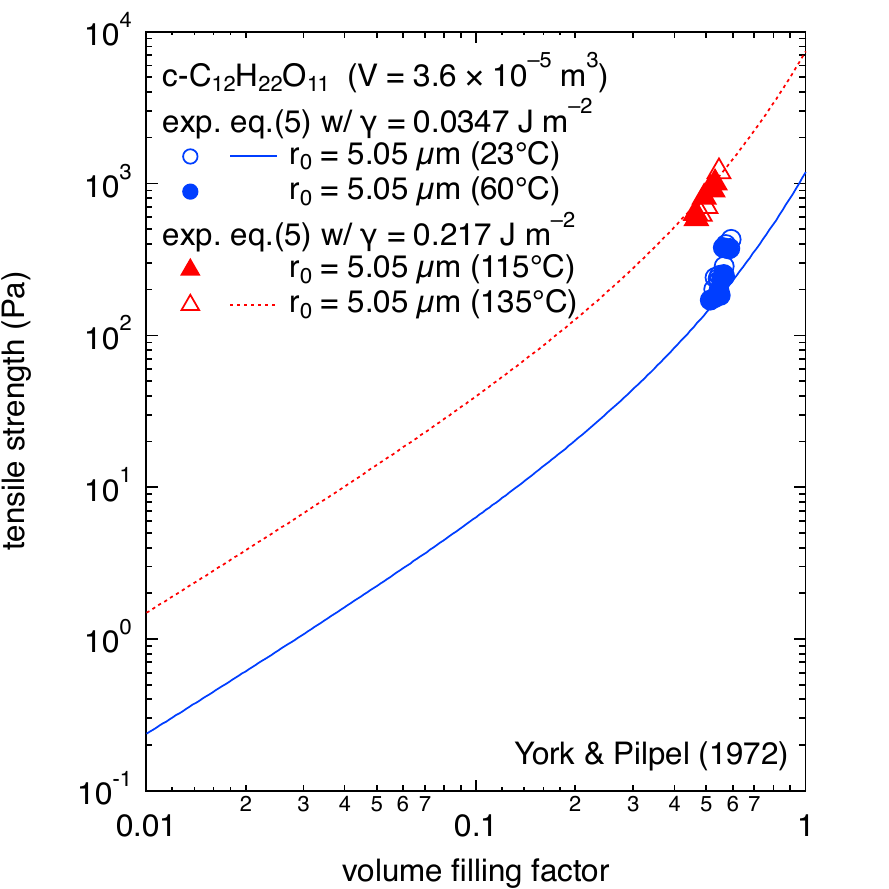}\includegraphics[width=0.4\columnwidth]{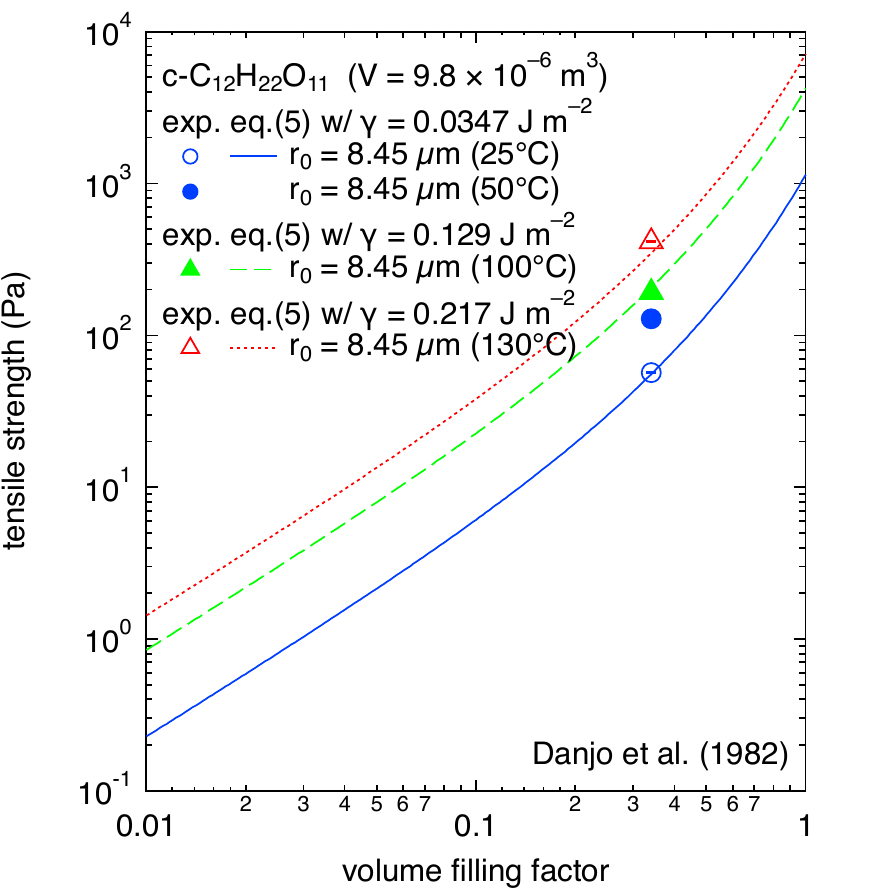}\\
\includegraphics[width=0.4\columnwidth]{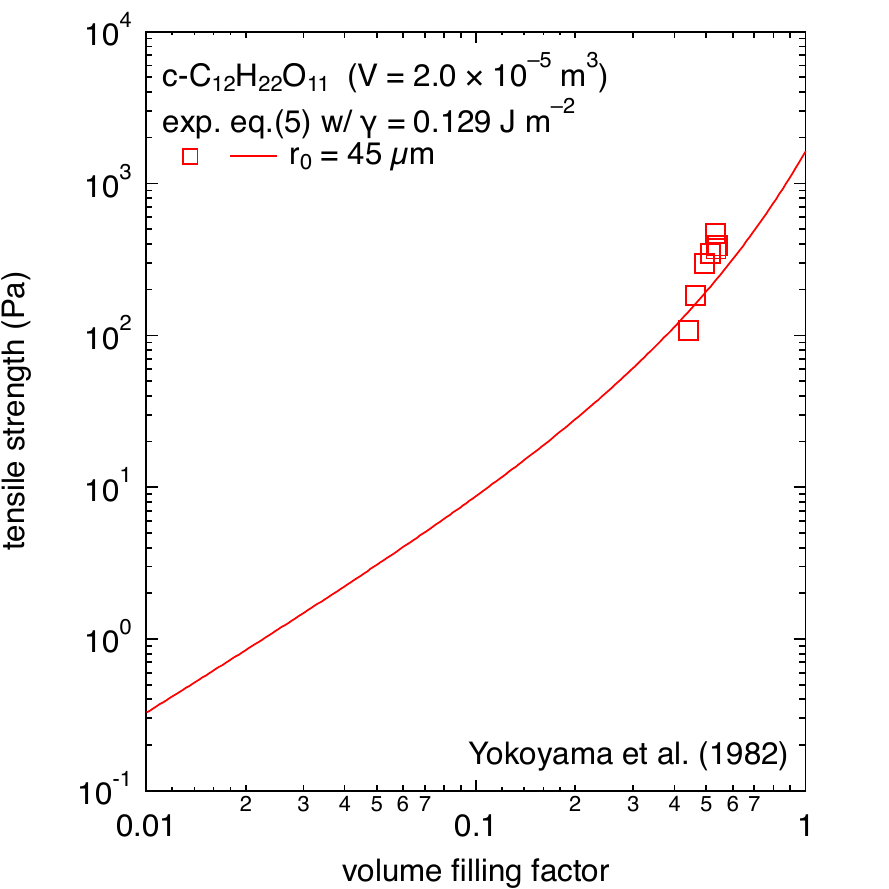}\includegraphics[width=0.4\columnwidth]{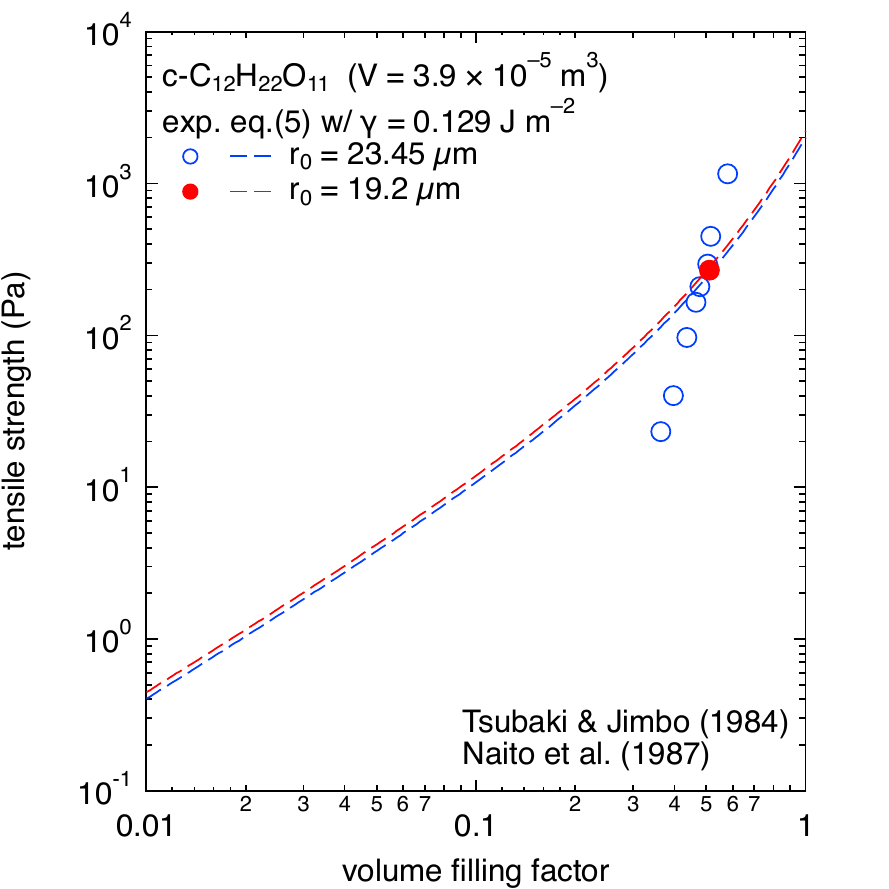}\\
\includegraphics[width=0.4\columnwidth]{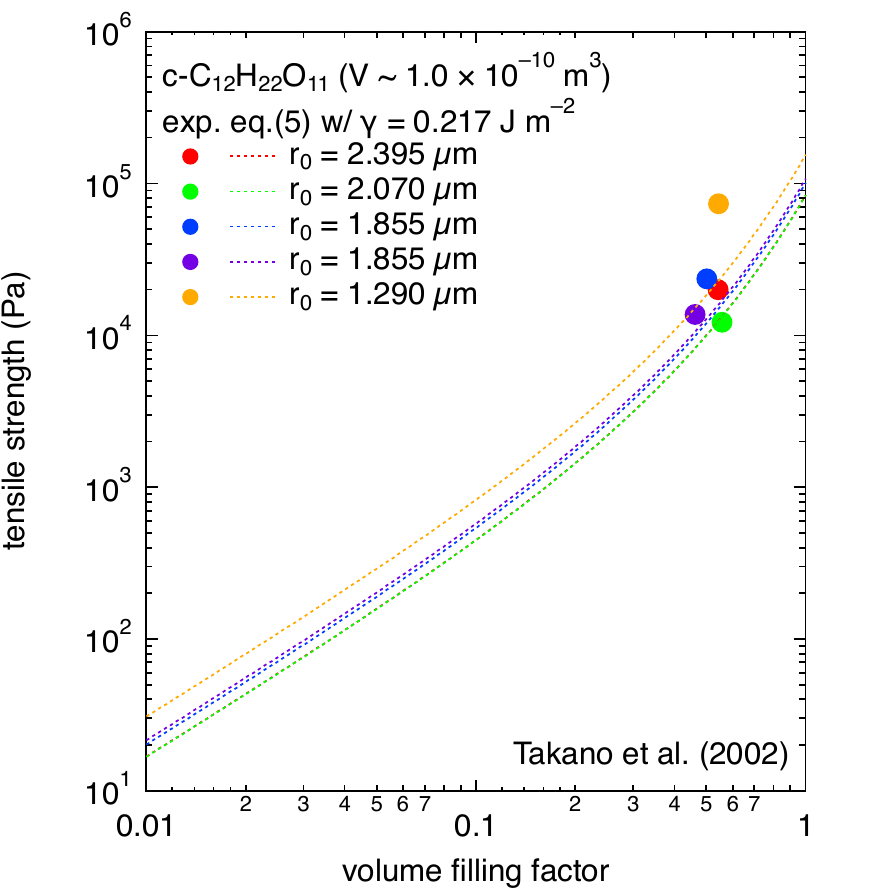}\includegraphics[width=0.4\columnwidth]{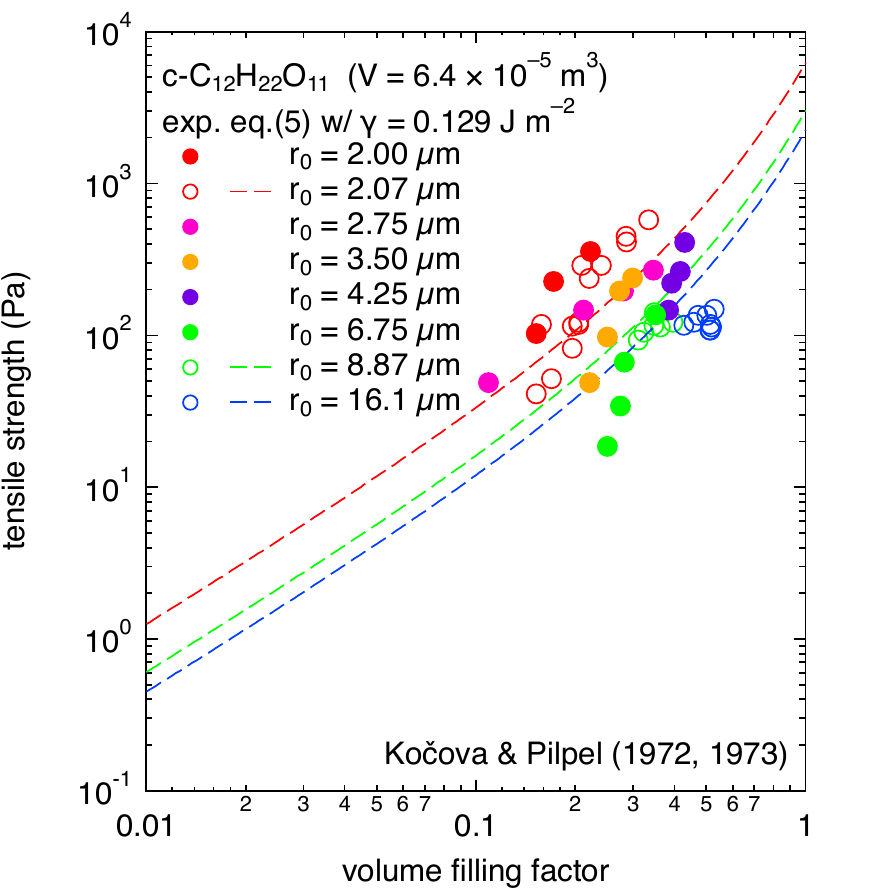}
\caption{A comparison of tensile strength for granular materials of polydisperse irregularly shaped lactose monomers between laboratory experiments in air by \citet[][left top]{york-pilpel1972}, \citet*[][right top]{danjo-et-al1982}; \citet*[][left middle]{yokoyama-et-al1982}; \citet{tsubaki-jimbo1984} and \citet[][right middle]{naito-et-al1987}, \citet[][left bottom]{takano-et-al2002}, and \citet{kocova-pilpel1972}, \citet[][right bottom]{kocova-pilpel1973} and our formula. The open and filled circles: experimental data for lactose powder beds at room temperature; the open and filled triangles: experimental data for lactose powder beds at slightly elevated temperature; the solid lines: equation~(\ref{eq:JKR}) with $\gamma = 0.0347~\mathrm{J~m^{-2}}$; the dotted line: equation~(\ref{eq:JKR}) with $\gamma = 0.217~\mathrm{J~m^{-2}}$.
\label{fig:experiments-lactose}}
\end{figure}

To the best of our knowledge, astronomically relevant carbonaceous matter has not been utilized for tensile strength measurements of dust aggregates.
Accordingly, we shall substitute lactose for astronomical organic matter, by taking into account the availability of tensile strength measurements for granular materials of lactose in air.
The surface energy of crystalline $\alpha$-lactose with a lack of surface contamination was measured to be $\gamma =0. 217$ and $0.129~\mathrm{J~m^{-2}}$ for crystalline $\alpha$-lactose anhydrous and crystalline $\alpha$-lactose monohydrate, respectively \citep*{traini-et-al2008,das-et-al2009,das-et-al2010,jones-et-al2012}.
It is well-known that lactose is hydrophilic and the surface of lactose at room temperature in air is easily covered by adsorbed water molecules, which reduce the surface energy.
The surface energy of crystalline $\alpha$-lactose monohydrate determined by \citet{sindel-zimmermann2001} using atomic force microscopy (AFM) was $\gamma = 0.0347~\mathrm{J~m^{-2}}$ at room temperature in air\footnote{\citet{sindel-zimmermann2001} measured a pull-off force $F$ of lactose surfaces using a lactose particle as a tip of their AFM cantilever. Since the pull-off force $F_\mathrm{ad}$ and the radius $R_\mathrm{tip}$ of the tip were determined to be $F_\mathrm{ad} = 5.0 \pm 3.06~\mathrm{nN}$ and $R_\mathrm{tip} = 15.3~\mathrm{nm}$, respectively, we obtain $\gamma = 0.0347 \pm 0.0212~\mathrm{J~m^{-2}}$ for the surface energy of $\alpha$-lactose monohydrate at room temperature in air. 
We are aware that \citet{zhang-et-al2006} derived $\gamma = 0.0233 \pm 0.0023~\mathrm{J~m^{-2}}$ for $\alpha$-lactose monohydrate from their AFM measurements with a silica tip at room temperature in air. Here, they assumed $\gamma = 0.042~\mathrm{J~m^{-2}}$ for silica to derive the surface energy of lactose from their measurements of $F_\mathrm{ad} = 6.34 \pm 0.35~\mathrm{nN}$ with $R_\mathrm{tip} = 21.6 \pm 0.6~\mathrm{nm}$.
It should be, however, noted that the surface energy of silica is strongly environmental dependent at room temperature in air and thus the assumption of $\gamma = 0.042~\mathrm{J~m^{-2}}$ for their silica tip cannot be justified \citep{kimura-et-al2015}.
Using an AFM from the same manufacture as \citet{zhang-et-al2006}, \citet{berard-et-al2002} obtained $F_\mathrm{ad} = 31~\mathrm{nN}$ for $\alpha$-lactose monohydrate at room temperature in air with a silica cantilever of $R_\mathrm{tip} = 20~\mathrm{nm}$.
The large discrepancy between the two results of pull-off force cannot be accounted for by the difference in the size of the tips, but most probably by the difference in the surface energy of the tips.}.

\citet{york-pilpel1972} studied experimentally how the tensile strength of crystalline $\alpha$-lactose monohydrate powders with $r_0 = 5.05~\micron$ varies with temperature of the powders.
Their results\footnote{The tensile strength of crystalline $\alpha$-lactose monohydrate powders was degraded at $180^\circ\mathrm{C}$, which may be attributed to thermal degradation of lactose, because pyrolysis of lactose takes place between $150^\circ\mathrm{C}$ and $200^\circ\mathrm{C}$ \citep{hohno-adachi1982}. Therefore, we shall disregard their experimental results on the tensile strength of $\alpha$-lactose powders obtained at temperatures higher than $150^\circ\mathrm{C}$.} show that the tensile strength of the powders is constant in the range of temperatire from $23$ to $60^\circ\mathrm{C}$, and elevated from $90$ to $160^\circ\mathrm{C}$.
The left top panel of Fig.~\ref{fig:experiments-lactose} shows their measurements on the tensile strength of polydisperse irregularly shaped lactose powders and equation~(\ref{eq:JKR}) with $\gamma = 0.0347~\mathrm{J~m^{-2}}$ (the solid line) and $\gamma = 0.217~\mathrm{J~m^{-2}}$ (the dotted line) as a function of the volume filling factor at room temperature as well as at elevated temperatures.
The experimental data at temperatures of $115$ and $135^\circ\mathrm{C}$ are situated on the dotted line of equation~(\ref{eq:JKR}) with $\gamma = 0.217~\mathrm{J~m^{-2}}$, consistent with thermal dehydration of monohydrate lactose around $120^\circ\mathrm{C}$ \citep[see][]{danjo-et-al1982}.

\citet{danjo-et-al1982} confirmed the temperature effect on the tensile strength of polydisperse irregularly shaped crystalline $\alpha$-lactose monohydrate powders with $r_0 = 8.45~\micron$ in air.
Their data shown in the right top panel of Fig.~\ref{fig:experiments-lactose} and in harmony with equation~(\ref{eq:JKR}) are obtained after they attained the selected temperatures in 30 minutes and then kept the temperatures for 4.5~h. 

\citet{yokoyama-et-al1982} prepared a powder bed of polydisperse lactose particles with $r_0 = 45~\micron$ by pressing the sample until the thickness of the powder bed reaches $10~\mathrm{mm}$.
The authors did not describe which form of lactose was used in their experiments, while we assume crystalline $\alpha$-lactose monohydrate that is the most common form of lactose at room temperature in air \citep{carpin-et-al2016}.
Since the compressed powder beds remained compact for 10~minutes as reported by the authors, we consider that the effect of adsorbed water molecules on the cohesion of particles is minimized for their samples.
Therefore, we may regard $\gamma = 0.129~\mathrm{J~m^{-2}}$ as the most appropriate value of surface energy for their powder beds, when comparing equation~(\ref{eq:JKR}) with their experimental data obtained at room temperature in air.
The left middle panel of Fig.~\ref{fig:experiments-lactose} shows that equation~(\ref{eq:JKR}) with $\gamma = 0.129~\mathrm{J~m^{-2}}$ reasonably reproduces their experimental data.

\citet{tsubaki-jimbo1984} presented their experimental data for the tensile strength of polydisperse $\alpha$-lactose monohydrate powders with $r_0 = 23.45~\micron$ measured at room temperature in air.
They pressed their powder beds prior to measurements and found that the tensile strength increases with the pre-compressive stress, which controls the volume filling factor of the powder beds.
\citet{naito-et-al1987} measured the tensile strength of polydisperse irregularly shaped $\alpha$-lactose monohydrate powders with $r_0 = 19.2~\micron$ at a temperature of $20^\circ\mathrm{C}$ and a relative humidity of 50\%.
While the former prepared their powder beds by compression in the range of $0.5 \times {10}^{3}$--$1 \times {10}^{5}~\mathrm{Pa}$, the latter $1.55 \times {10}^{3}$--$11.5 \times {10}^{3}~\mathrm{Pa}$.
Their powder beds were prepared in the same volume and thus their results are plotted together in the right middle panel of Fig.~\ref{fig:experiments-lactose}.
The coincidence of equation~(\ref{eq:JKR}) with their results of tensile strength is fairly good, although the compression during the sample preparation seems to affect the tensile strength.

The tensile strength of nearly spherical compact aggregates composed of polydisperse crystalline lactose particles was measured by \citet{takano-et-al2002} in the ranges of volume from $V = 2.7 \times {10}^{-11}$ to $2.6 \times {10}^{-10}~\mathrm{m}^{3}$, depending on the radius of monomers.
Since a special type of dry granulation, referred to as pressure swing granulation (PSG), was utilized for milled $\alpha$-lactose particles to tightly agglomerate together, lactose particles are in contact without help of adsorbed water molecules on their surfaces.
Therefore, we may adopt $\gamma = 0.248~\mathrm{J~m^{-2}}$ for nearly spherical compact dust aggregates consisting of polydisperse crystalline lactose monomers produced by the PSG method.
The experimental values of tensile strength are scattered around equation~(\ref{eq:JKR}), but the result for the aggregates consisting of the smallest monomers with $r_0 = 1.29~\micron$ greatly exceeds the tensile strength expected from equation~(\ref{eq:JKR}).

\citet{kocova-pilpel1972,kocova-pilpel1973} used irregularly shaped polydisperse powders of crystalline $\alpha$-lactose monohydrate with $r_0 = 2.00$, $2.07$, $2.75$, $3.50$, $4.25$, $6.75$, $8.87$, $16.1~\micron$ for their measurements of tensile strength at room temperature in air.
Since they performed their measurement with a dehumidifier after the powders were dried at a temperature of $105^\circ\mathrm{C}$, we may assume $\gamma = 0.129~\mathrm{J~m^{-2}}$ for their dried $\alpha$-lactose powders.
While their results are scattered around equation~(\ref{eq:JKR}), irrespective of monomer size, there does not seem to show a clear discrepancy between equation~(\ref{eq:JKR}) and their results.

\subsection{Astronomical observations}

\subsubsection{Cometary dust}

\begin{figure}
\centering
\includegraphics[width=0.4\columnwidth]{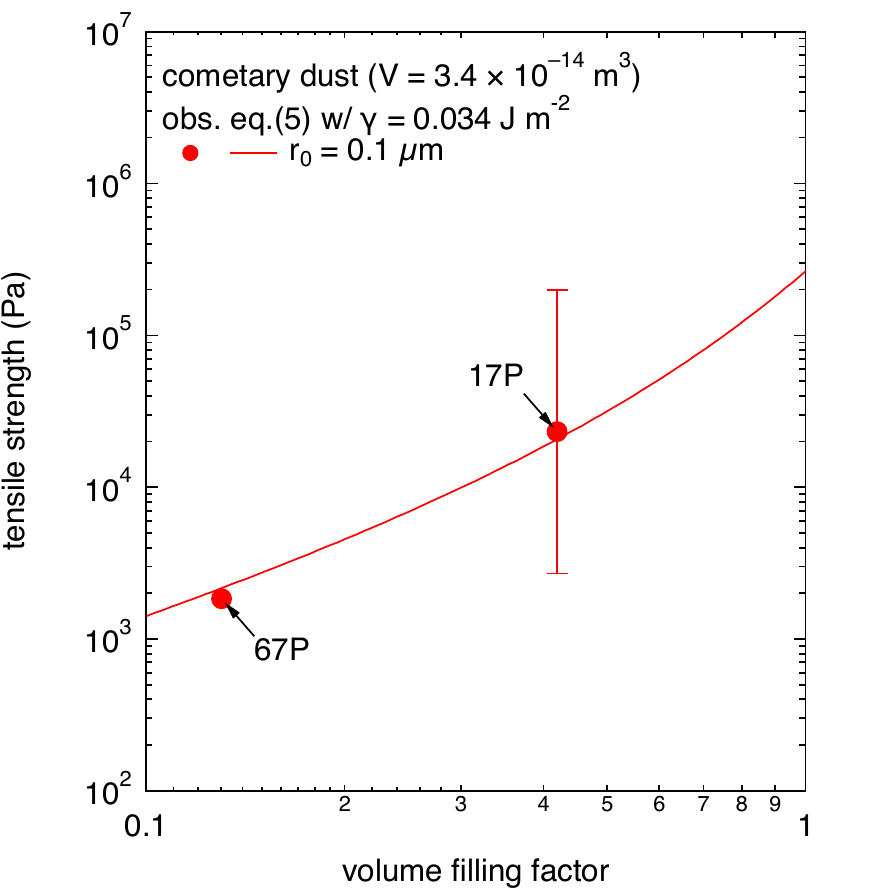}\includegraphics[width=0.4\columnwidth]{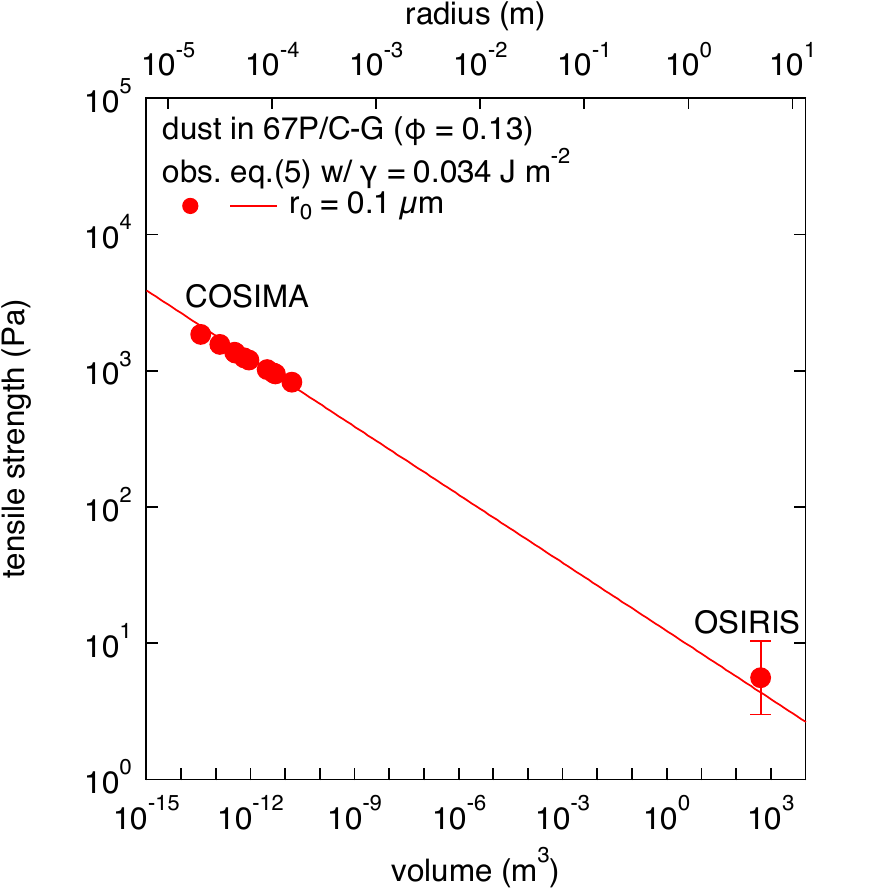}
\caption{A comparison of tensile strength between cometary dust (the filled circles) and porous dust aggregates of monodisperse spherical monomers in our formula with $r_0 = 0.1~\micron$ and $\gamma = 0.034~\mathrm{J~m^{-2}}$ (the solid lines). Left: dust particles with radius $R = 20~\micron$ in the comae of 17P/Holmes and 67P/Churyumov-Gerasimenko. Right: dust particles with volume filling factor $\phi = 0.13$ in the coma and on the surface of 67P/Churyumov-Gerasimenko observed by the COSIMA and OSIRIS instruments onboard Rosetta.
\label{fig:Rosetta}}
\end{figure}

Thanks to an explosion on Comet 17P/Holmes, \citet{reach-et-al2010} were able to estimate the tensile strength of dust particles ejected from the surface of the comet.
They found that the tensile strength of $\sigma = 23.2^{+176.8}_{-20.5}~\mathrm{kPa}$ is required to release dust particles with radius $R = 20^{+180}_{-18}~\micron$.
\citet{hornung-et-al2016} estimated the tensile strength of dust particles ejected from 67P/C-G using COSISCOPE images of dust aggregates in the size range of $R = 20$--$155~\micron$ collected by the Rosetta/COSIMA instrument.
COSIMA/Rosetta images and mass spectra of dust aggregates in the coma of 67P/C-G show that the surface chemistry of the aggregates is consistent with carbonization of organic matter characterized by the surface energy of $\gamma = 0.034~\mathrm{J~m^{-2}}$ \citep{kimura-et-al2020}.
The left panel of Fig.~\ref{fig:Rosetta} shows that the tensile strength of dust particles with $R = 20~\micron$ is well reproduced by equation~(\ref{eq:JKR}) with $r_0 = 0.1~\micron$ and $\gamma = 0.034~\mathrm{J~m^{-2}}$.

\citet{groussin-et-al2015} derived the tensile strength of overhangs on the surface of 67P/C-G to be $\sigma = 5.6^{+9.4}_{-2.6}~\mathrm{Pa}$ from Rosetta/OSIRIS images by assuming the length of $10~\mathrm{m}$ and the hight of $5~\mathrm{m}$.
Because the size of boulders located at the feet of these overhangs is approximately $10~\mathrm{m}$, we take $10~\mathrm{m}$ as the width of the overhangs to estimate the volume of the overhangs.
OSIRIS images revealed that photometric and spectrophotometric data are best explained by the Hapke's reflectance model if the volume filling factor of its surface is $\phi = 0.13$ \citep{fornasier-et-al2015}.
The right panel of Fig.~\ref{fig:Rosetta} depicts the tensile strength of dust aggregates for the surface of comet 67P/C-G obtained by COSISCOPE images of dust particles and OSIRIS images of overhangs.
This clearly shows the volume effect on the tensile strength for cometary dust aggregates, which is accounted for by equation~(\ref{eq:JKR}) with $r_0 = 0.1~\micron$ and $\gamma = 0.034~\mathrm{J~m^{-2}}$.

\subsubsection{Meteor showers}

\begin{figure}
\centering
\includegraphics[width=0.4\columnwidth]{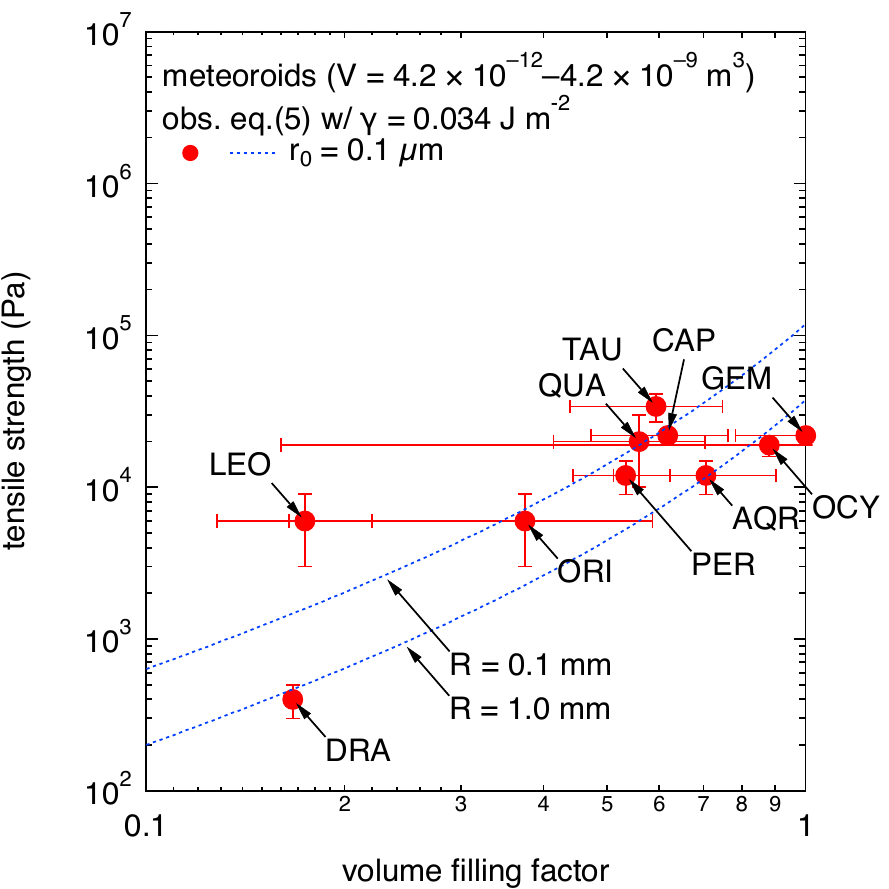}\includegraphics[width=0.4\columnwidth]{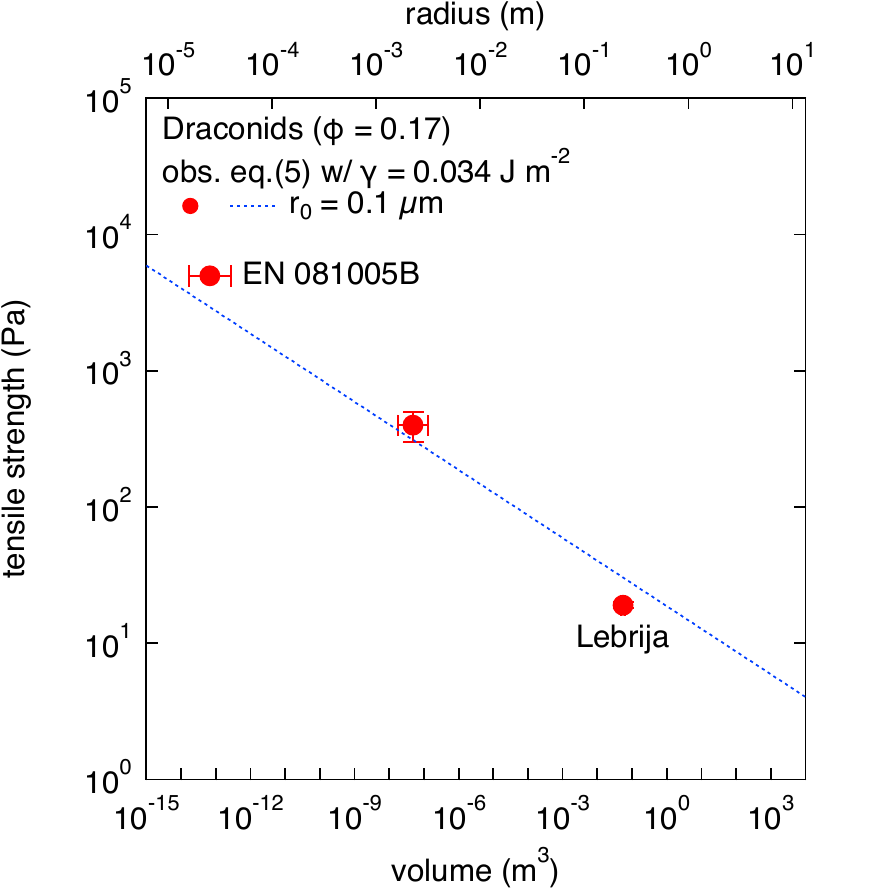}
\caption{A comparison of tensile strength between cometary meteoroids based on meteor observations (the filled circles) and dust aggregates of monodisperse spherical monomers based on our formula: equation~(\ref{eq:JKR}) with $r_0 = 0.1~\micron$ and $\gamma = 0.034~\mathrm{J~m^{-2}}$.
Left: the blue-dotted line: equation~(\ref{eq:JKR}) with aggregate radii $R = 0.1$ and $1.0~\mathrm{mm}$; AQR: $\delta$-Aquarids; CAP :$\alpha$-Capricornids; DRA: Draconids (Giacobinids) ; GEM: Geminids; LEO: Leonids; OCY: $o$-Cygnids; ORI: Orionids; PER: Perseids; QUA: Quadrantids; TAU: Taurids. Right: the blue-dotted line: equation~(\ref{eq:JKR}) with volume filling factor $\phi = 0.17$; the filled circles: Draconid meteor observations inclusive of EN 081005B and Lebrija Draconid fireballs.
\label{fig:observations}}
\end{figure}
\citet{trigorodriguez-llorca2006} determined the tensile strength of cometary meteoroids based on ground-based observations of meteor showers, which suggests an increase in the strength with the density of meteoroids.
Since they claim that a typical radius of meteoroids presented in their paper is $r < 5~\mathrm{mm}$, we consider two apparent radii of $R = 0.1~\mathrm{mm}$ ($V = 4.2 \times {10}^{-12}~\mathrm{m^3}$) and $R = 1.0~\mathrm{mm}$ ($V = 4.2 \times {10}^{-9}~\mathrm{m^3}$).
On the basis of a model for quasi-continuous fragmentation of meteoroids, \citet{babadzhanov-kokhirova2008} determined the porosities of meteoroids, which are converted into the volume filling factors $\phi$ in our study.
The left panel of Fig.~\ref{fig:observations} comapres the tensile strength of cometary meteoroids with various $\phi$ values derived from ground-based observations of meteor showers to equation~(\ref{eq:JKR}) with $r_0 = 0.1~\micron$ and $\gamma = 0.034~\mathrm{J~m^{-2}}$. 
We find that the tensile strengths of cometary meteoroids are scattered along equation~(\ref{eq:JKR}) with a radius of $R = 0.1$ and $1.0~\mathrm{mm}$.

\citet*{borovicka-et-al2007} derived the tensile strength of $\sigma = 5~\mathrm{kPa}$ for the EN 081005B Draconid fireball with radius $R = 0.025^{+0.015}_{-0.009}~\mathrm{mm}$ from its light curve photographed by Super-Schmidt cameras, similar to the precedent study by \citet{trigorodriguez-llorca2006}.
\citet{trigorodriguez-llorca2006} estimated the tensile strengths of meteoroids using various observed meteor data inclusive of \citet{fujiwara-et-al2001}, from which we obtain $R = 2.25^{+0.90}_{-0.64}~\mathrm{mm}$ for Draconids by assuming a density of $300~\mathrm{kg~m^{-3}}$.
\citet{madiedo-et-al2013} derived the tensile strength of $\sigma = 19 \pm 1~\mathrm{Pa}$ for an extraordinary bright Draconid fireball `Lebrija' with radius $R = 0.23~\mathrm{m}$ from their observations at multiple stations.
The right panel of Fig.~\ref{fig:observations} shows the tensile strengths of typical Draconids with $R \approx 0.02$--$2.0~\mathrm{mm}$ and an extraordinal Draconid fireball of $R \approx 0.2~\mathrm{m}$, and equation~(\ref{eq:JKR}) with $r_0 = 0.1~\micron$, $\gamma = 0.034~\mathrm{J~m^{-2}}$, and $\phi = 0.17$.
A reasonable fit of equation~(\ref{eq:JKR}) to observations of Draconids justifies the validity of equation~(\ref{eq:JKR}) to describe the volume effect on the tensile strength of dust aggregates, although the tensile strengths of the EN 081005B and Lebrija fireballs lie slightly above and below equation~(\ref{eq:JKR}), respectively.

\section{Discussion}

We find that equation~(\ref{eq:JKR}) is capable of reproducing the dependences of tensile strength on the volume filling factor, irrespective of monomer's composition, size, crystallinity, and surface chemistry.
However, we admit that the tensile strength of compact dust aggregates consisting of the smallest $\mathrm{sicastar}\textsuperscript{\textregistered}$ spheres with $r_0 = 0.15~\micron$ measured in air by \citet{gundlach-et-al2018} exceptionally exceeds the value expected by equation~(\ref{eq:JKR}) with $\gamma = 0.150~\mathrm{J~m^{-2}}$ (see the right top panel of Fig.~\ref{fig:experiments-sphere}).
It is worthwhile noting that amorphous silica particles at room temperature in air are known to swell up by adsorption of water molecules, owing to its hydrophilic nature \citep{vigil-et-al1994,zhuravlev2000}.
\citet{steinpilz-et-al2019} estimated the thickness $\Delta r_0$ of water layers on the surface of $\mathrm{sicastar}\textsuperscript{\textregistered}$ to be $\Delta r_0 = (25.3 \pm 4.0)~\mathrm{nm}$ at atmospheric conditions.
It turned out that the smaller the size of monomers is, the stronger the effect of adsorbed water on the volume filling factor of dust aggregates is (see Appendix~\ref{appendix:water}).
If we assume the same thickness of water molecules for $\mathrm{sicastar}\textsuperscript{\textregistered}$ spheres with $r_0 = 0.15~\micron$, then we find that the volume filling factor of compact dust aggregates was $\phi=0.54$, instead of $\phi=0.44$.
Therefore, the deviation of equation~(\ref{eq:JKR}) from the experimental data on the tensile strength of compact aggregates with $r_0 = 0.15~\micron$ could at least partly be attributed to underestimation of volume filling factors due to water adsorption in laboratory experiments.

Another clear mismatch between experiments and our formula is the tensile strength of PSG lactose granules consisting of polydisperse monomers with $r_0 = 1.29~\micron$ produced by \citet{takano-et-al2002} (see the left bottom panel of Fig.~\ref{fig:experiments-lactose}).
Their SEM (Scanning Electron Microscope) images of PSG granules show that the compactness of a granule with $r_0 = 1.29~\micron$ is distinct from that with the other monomers' radii.
Because the granule with $r_0 = 1.29~\micron$ appears as a single compact sphere in the SEM image, the volume filling factor of the granule could be as high as $\phi \approx 0.78$ \citep[cf.][]{beck-volpert2003}.
While we cannot rule out model limitations, an underestimation of the volume filling factor for the specific granule may be a remedy for the discrepancy between the experiments and the model.

One may notice in the right bottom panel of Fig.~\ref{fig:DEMsimulations} that equation~(\ref{eq:JKR}) predicts the tensile strength of dust aggregates consisting of monodisperse spherical monomers a factor of two larger than the results of DEM simulations by \citet{seizinger-et-al2013}.
On closer inspection, however, the right top and left bottom panels of Fig.~\ref{fig:DEMsimulations} shows that the tensile strength determined by \citet{seizinger-et-al2013} is smaller by a factor of two compared to \citet{tatsuuma-et-al2019}, although they both rely on the JKR theory for dust aggregates with $r_0 = 0.6~\micron$ and $\gamma = 0.02~\mathrm{J~m^{-2}}$.
One of the noticeable differences in their numerical simulations is a time step for integration; the former uses $1$--$3 \times {10}^{-10}~\mathrm{sec}$, while the latter $1.9 \times {10}^{-11}~\mathrm{sec}$.
It is most likely that the larger the time steps in DEM simulations are, the higher the possibility of overlooking the maximum tensile stress in the simulations is.
Therefore, we conclude that DEM simulations by \citet{seizinger-et-al2013} may be underestimated by a factor of 2, owing to the use of large time steps in their simulations.

We find that equation~(\ref{eq:JKR}) slightly underestimates and overestimates the tensile strength of the Draconid fireballs `EN 081005B' and `Lebrija', respectively (the right panel of Fig.~\ref{fig:observations}), if we use the volume filling factor of $\phi = 0.17$, according to \citet{babadzhanov-kokhirova2008}.
While \citet{babadzhanov-kokhirova2008} estimated the porosity (i.e. the volume filling factor) of Draconids based on the density of $300~\mathrm{kg~m^{-3}}$, \citet{madiedo-et-al2013} suggested a lower density of $100~\mathrm{kg~m^{-3}}$ for the Lebrija fireball.
Because the tensile strength increases with the density, in other words, the volume filling factor as expressed in equation~(\ref{eq:JKR}), it is reasonable to attribute the deviation of the Lebrija fireball from our prediction to the low density of the Lebrija fireball.
As a result, we cannot rule out the possibility that the density of meteoroids decrease with radius, as inferred for dust particles in the coma of comet 1P/Halley from photopolarimetric properties of the particles \citep*[see][]{lamy-et-al1987}.
Consequently, equation~(\ref{eq:JKR}) is still valid for estimating the tensile strength of cometary meteoroids, if we assume $r_0 = 0.1~\micron$ and $\gamma = 0.034~\mathrm{J~m^{-2}}$.
We should, however, remind the reader that a comprehensive analysis of meteor data obtained at multiple stations will certainly provide valuable information on the mineralogical and morphological properties of cometary meteoroids.

\begin{figure}
\centering
\includegraphics[width=0.4\columnwidth]{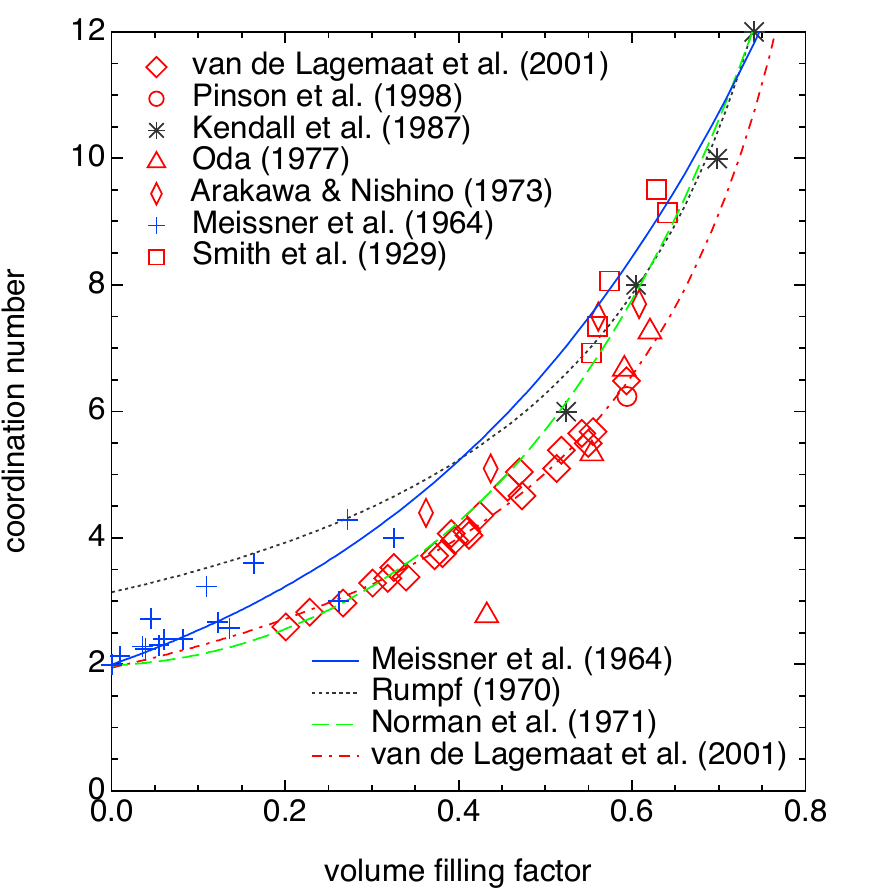}
\caption{Models of a prescription for the relationship between the coordination number $n_\mathrm{c}$ and the volume filling factor $\phi$, compared to data on the $n_\mathrm{c}$--$\phi$ relationship from \citet*{smith-et-al1929}, \citet{meissner-et-al1964}, \citet{arakawa-nishino1973}, \citet{oda1977}, \citet{kendall-et-al1987}, \citet{pinson-et-al1998}, and \citet*{vandelagemaat-et-al2001}. Solid line: \citet{meissner-et-al1964}; dotted line: \citet{rumpf1970}; dashed line: \citet*{norman-et-al1971}; dash-dotted line: \citet{vandelagemaat-et-al2001}.
\label{fig:nc-phi-relation}}
\end{figure}
We have demonstrated the validity of equation~(\ref{eq:JKR}), which incorporates equation~(\ref{eq:meissner-relation}), but there is room for improvement of equation~(\ref{eq:JKR}). 
Indeed, we cannot rule out the possibility that there is a better prescription for the relationship between the coordination number $n_\mathrm{c}$ and the volume filling factor $\phi$ of dust aggregates, compared with equation~(\ref{eq:meissner-relation}).
For example, the classical model of \citet{rumpf1970} suggests
\begin{eqnarray}
n_\mathrm{c} & = & \frac{\pi}{1-\phi} ,
\label{eq:rumpf-relation}
\end{eqnarray}
while \citet{gundlach-et-al2018} considered that a reasonable prescription for the $n_\mathrm{c}$-$\phi$ relationship of dust aggregates is given by \citet{vandelagemaat-et-al2001}:
\begin{eqnarray}
n_\mathrm{c} & = & \frac{c_1}{1-\phi} - c_2 ,
\label{eq:vandelagemaat-relation}
\end{eqnarray}
with $c_1=3.08$ and $c_2=1.13$.
\citet{norman-et-al1971} proposed an extension of equation~(\ref{eq:meissner-relation}):
\begin{eqnarray}
n_\mathrm{c} & = & c_1 \exp \left({d_1 \phi}\right) - c_2 \exp \left({d_2 \phi}\right) ,
\label{eq:norman-relation}
\end{eqnarray}
where $c_1 = 1.126$, $c_2 = 0.860$, $d_1 = 3.196$, and $d_2 = - 3.50$.
Figure~\ref{fig:nc-phi-relation} depicts these models for the $n_\mathrm{c}$-$\phi$ relationships together with the data for specific structures of granular matters.
It turned out that equation~(\ref{eq:meissner-relation}) gives the highest $n_\mathrm{c}$ values in the range of $\phi \approx 0.4$--$0.7$, compared with the other models.\footnote{Note that these models given in Eqs.~(\ref{eq:rumpf-relation})--(\ref{eq:norman-relation}) are not all, but merely three examples; 
There are plenty of formulae that provide a prescription for the relationship between the coordination number $n_\mathrm{c}$ and the volume filling factor $\phi$ \citep[see][]{vanantwerpen-et-al2010}.}
This could partly explain the reason that equation~(\ref{eq:JKR}) tends to slightly overestimate the tensile strength of dust aggregates in the range of $\phi = 0.41$--$0.66$ measured by \citet{blum-schraepler2004} and \citet{blum-et-al2006}.
Therefore, we expect that a better choice of the $n_\mathrm{c}$-$\phi$ relationship would improve a theoretical prediction for the tensile strength of porous dust aggregates.

\begin{figure}
\centering
\includegraphics[width=0.4\columnwidth]{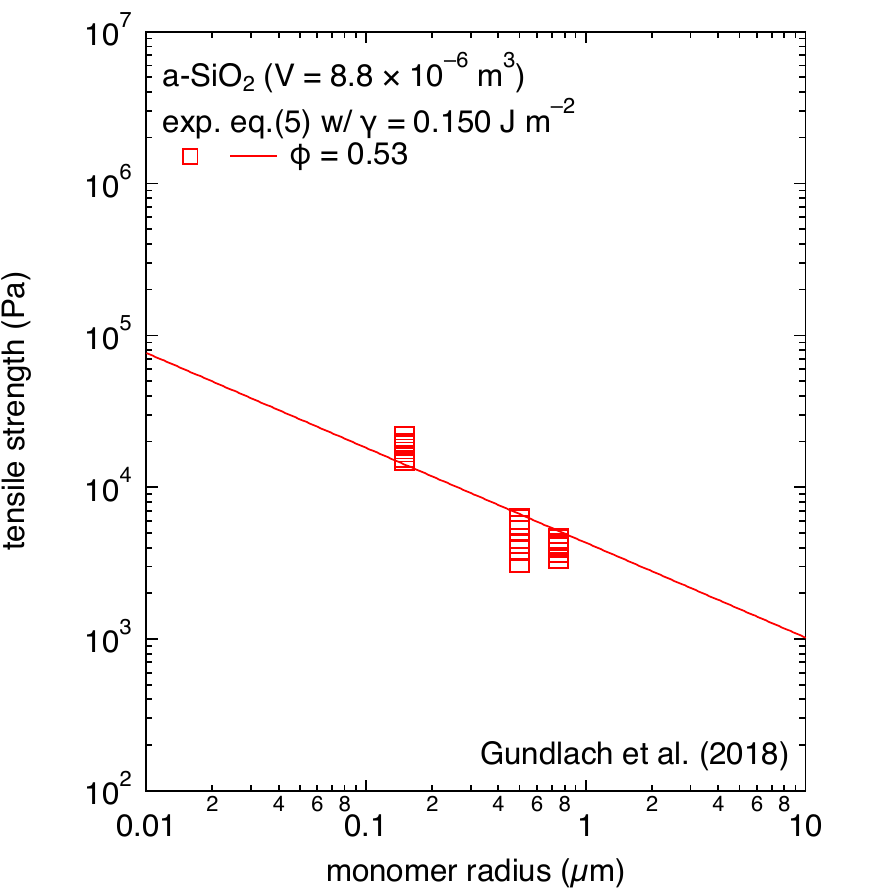}
\caption{The dependence of tensile strength on the radius of monomers for dust aggregates expected from laboratory experiments and our formula. The open squares: experimental results by \citet{gundlach-et-al2018}.
\label{fig:monomer-radius-effect}}
\end{figure}
There has been no consensus about the dependence of tensile strength $\sigma$ on the radius $r_0$ of monomers among models for the tensile strength of dust aggregates: $\sigma \propto r_0^{-2.0}$ by \citet*{greenberg-et-al1995}; $\sigma \propto r_0^{-5/3}$ by \citet{wada-et-al2008}; $\sigma \propto r_0^{-1.0}$ by \citet{tatsuuma-et-al2019}.
It should be noted that these proportionalities are purely predictions by the respective models, but they have never been fully justified by experimental results up to date.
Our model implies $\sigma \propto r_0^{-1.0}$ as a model of \citet{tatsuuma-et-al2019}, if the volume of dust aggregates is proportional to the third power of $r_0$.
However, as far as the same volume of dust aggregates is concerned, we predict that the tensile strength of porous dust aggregates shows a weaker dependence of monomer radius as $\sigma \propto r_0^{3/m-1}$ (crudely $\sigma \propto r_0^{-0.5 \pm 0.1}$ for $m = 5$--$8$).
\citet{currier-schulson1982} presented their experimental results of $\sigma \propto r_0^{-0.5}$ for aggregates of polycrystalline water ice grains with $\phi = 0.999$, although the volume filling factor of $\phi = 0.999$ would lie beyond the applicability of our model.
Figure~\ref{fig:monomer-radius-effect} depicts that the dependence of tensile strength on the radius of monomers measured for the same volume of dust aggregates by \citet{gundlach-et-al2018} is consistent with equation~(\ref{eq:JKR}).
Our success in reproducing experimental and numerical results of tensile strength, irrespective of the composition and the size of monomers as well as the volume of the aggregates, presented in Sec~\ref{sec:results} has given grounds for the proportionality of $\sigma \propto r_0^{3/m-1}$.

\begin{figure}
\centering
\includegraphics[width=0.4\columnwidth]{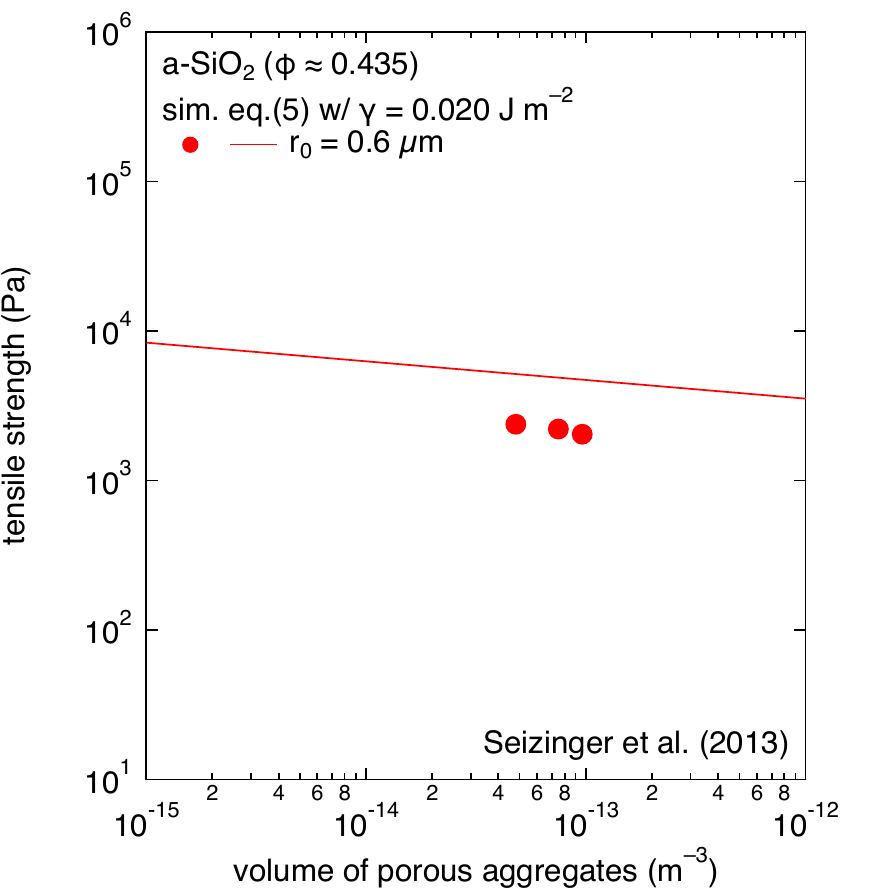}\includegraphics[width=0.4\columnwidth]{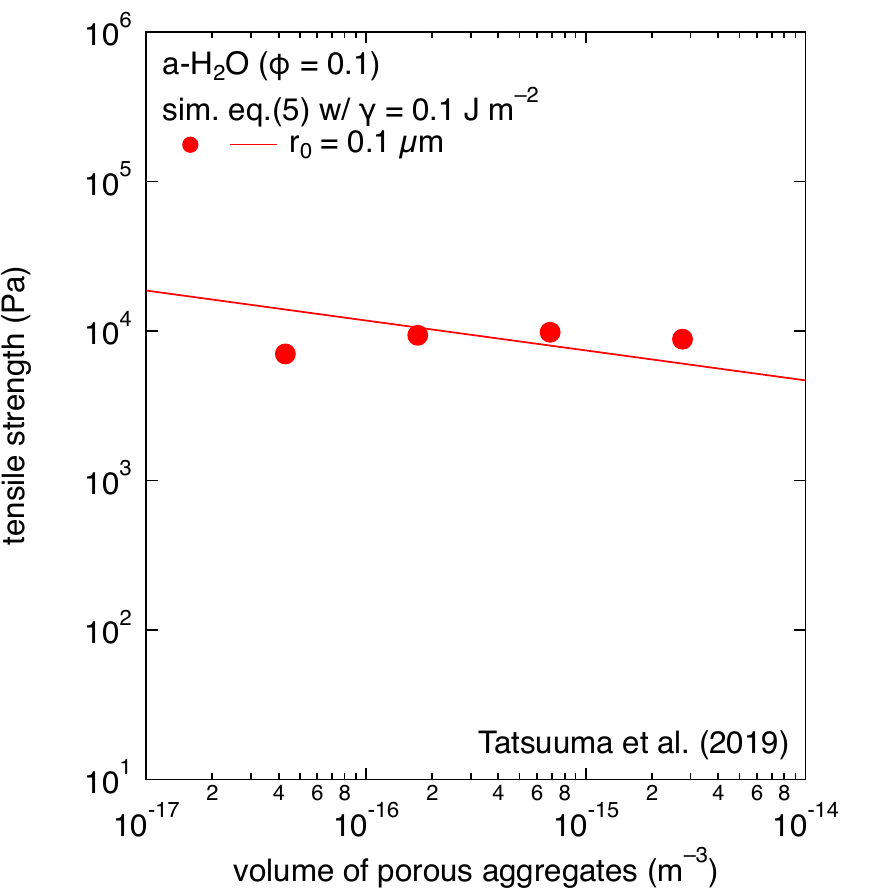}\\
\includegraphics[width=0.4\columnwidth]{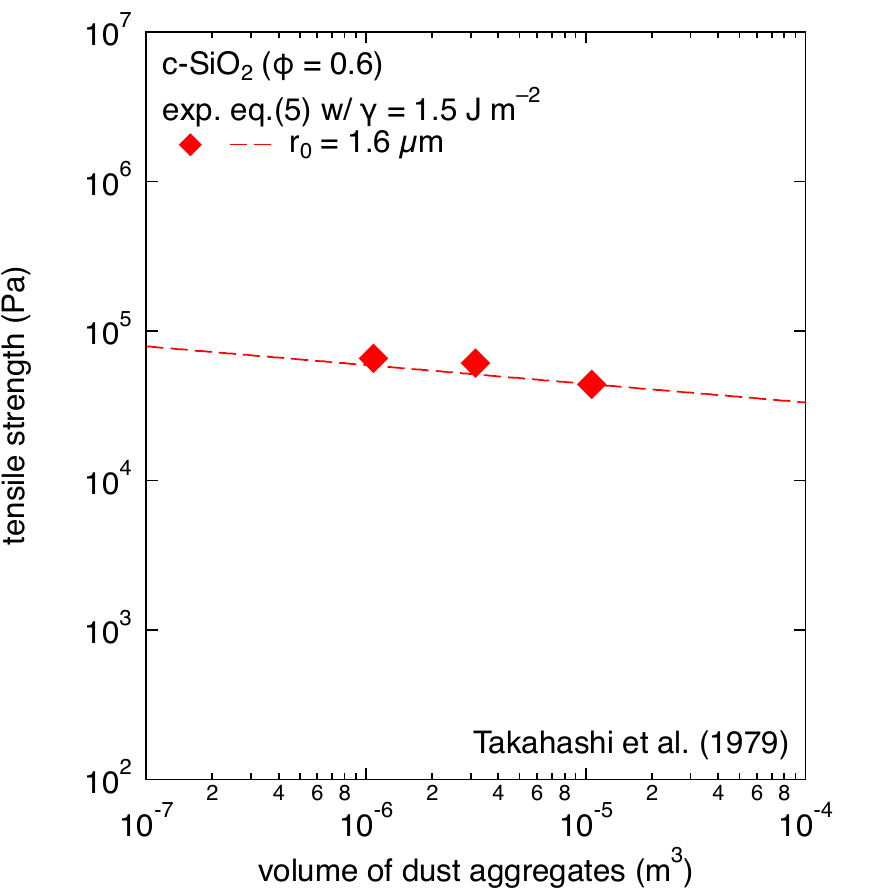}
\caption{The volume effect of tensile strength for porous and compact dust aggregates expected from computer simulations, laboratory experiments, and our formula. Left top: numerical simulations by \citet{seizinger-et-al2013}; right top: numerical simulations by \citet{tatsuuma-et-al2019}; bottom: experimental results by \citet{takahashi-et-al1979}. 
\label{fig:volume-effect}}
\end{figure}
\citet{skorov-blum2012} proposed that the tensile strength $\sigma$ of dust aggregates increases with the volume of the aggregates, according to $\sigma \propto V^{2/15}$, up to millimeter sizes, and then decreases with the volume $V$ of the aggregates, according to $\sigma \propto V^{-2/9}$.
On the basis of numerical simulations, however, \citet{seizinger-et-al2013} and \citet{tatsuuma-et-al2019} concluded that the tensile strength of dust aggregates smaller than millimeter sizes does not depend on the volume of the aggregates\footnote{\citet{seizinger-et-al2013} considered the volume effect in the range of $V = 4.8$--$9.6 \times {10}^{-14}~\mathrm{m^{3}}$ ($40~\micron \times 40~\micron \times 30~\micron$--$40~\micron \times 40~\micron \times 60~\micron$) and \citet{tatsuuma-et-al2019} in the range of $V = 4.3 \times {10}^{-17}$--$2.7 \times {10}^{-15}~\mathrm{m^{3}}$ ($3.5~\micron \times 3.5~\micron \times 3.5~\micron$--$14~\micron \times 14~\micron \times 14~\micron$).}.
Our formula given in equation~(\ref{eq:JKR}) does not provide evidence for neither an increase in the tensile strength with the volume of small aggregates nor the volume independence of tensile strength.
The top panels of Fig.~\ref{fig:volume-effect} depicts the tensile strengths of dust aggregates numerically determined by \citet{seizinger-et-al2013} at $\phi \approx 0.435$ (left) and by \citet{tatsuuma-et-al2019} at $\phi = 0.1$ (right) as a function of the volume of their aggregates.
Here, one may notice that numerical results of \citet{seizinger-et-al2013} reveal a weak decline of tensile strength with the volume of their aggregates, while numerical results of \citet{tatsuuma-et-al2019} are scattered around equation~(\ref{eq:JKR}).
By taking into account the fact that the numerical results of \citet{seizinger-et-al2013} underestimated the tensile strength of dust aggregates by a factor of 2, their results are consistent with the volume effect of our formula given in equation~(\ref{eq:JKR}).
A lack of volume effects in the numerical results of \citet{tatsuuma-et-al2019} could be attributed to the small size of porous ($\phi = 0.1$) dust aggregates between $N = 2^{10}$--$2^{16}$ used in their simulations, because the size of the aggregates is on the same order as the maximum flaw size. 
Therefore, we anticipate that they would have also presented the volume effects, if larger volumes and higher volume filling factors were adopted in their simulations.
\citet{takahashi-et-al1979} have shown from their experiments that the tensile strength of silica powder beds with $\phi = 0.6$ gradually decreases with the volume of the powder beds, as expected by equation~(\ref{eq:JKR}) (see the bottom panel of Fig.~\ref{fig:volume-effect}).
Their results on the volume effect are again reproduced by equation~(\ref{eq:JKR}) with $\gamma = 1.5~\mathrm{J~m^{-2}}$, implying the establishment of siloxane bridges between monomers in the powder beds after intense compression.
Since cometary dust and meteoroids also exhibit volume effects in the results of in-situ and ground-based observations (see Figs.~\ref{fig:Rosetta} and \ref{fig:observations}), it is natural to consider that the tensile strength of dust aggregates gradually decreases with the volume of the aggregates, as expected from fracture mechanics (i.e. $\sigma \propto V^{-1/m}$).

As we have demonstrated throughout this paper using equation~(\ref{eq:JKR}), the volume effects may play a vital role in the predicted values of tensile strength, unless the Weibull modulus is large enough.
However, numerical results based on DEM simulations easily overlook this important effects as shown in the top panels of Fig.~\ref{fig:volume-effect}, due to a shortcoming of numerical simulations, which has a difficulty of dealing with a large span of volumes.
As a result, from a theoretical point of view, there is a great demand for the determination of the Weibull modulus for astronomically relevant materials by laboratory experiments.
Similarly important is a thorough inspection of a flaw size distribution in the laboratory, since the size distribution of flaws without a power law might violate the validity of equation~(\ref{eq:JKR}).
Therefore, we would like to encourage experimentalists to conduct their laboratory experiments with a wide size range of dust aggregates and to measure their tensile strengths and flaw-size distributions.

By taking into account uncertainties in the $n_\mathrm{c}$-$\phi$ relationship, the Weibull modulus, and the volume filling factor of dust aggregates consisting of swelling monomers in air, an analytical model of equation~(\ref{eq:JKR}) for the tensile strength of dust aggregates is capable of reproducing results of laboratory experiments and computer simulations. 
In addition, we have revealed that the tensile strength of dust aggregates consisting of submicrometer-sized monomers with $r_0 \approx 0.1~\micron$ in our model is consistent with observations of cometary dust and meteor showers.
In summary, we succeed in restoring the consensus that porous dust aggregates, which were incorporated into comets in the solar nebula, consist of solar nebular condensates with radius $r_0 \approx 0.1~\micron$.

\section*{Acknowledgements}

We would like to thank Akiko M. Nakamura, Misako Tatsuuma, and Josep M. Trigo-Rodriguez for profitable discussion on the tensile strength of porous dust aggregates from experimental, numerical, and observational points of view, respectively, Irina L. San Sebasti\'{a}n and J\"{u}rgen Blum for sharing their experimental data with us prior to publication, and  
an anonymous reviewer for her/his comments that helped us to improve the manuscript.
H.K. is grateful to JSPS's Grants-in-Aid for Scientific Research (KAKENHI \#19H05085).








\appendix

\section{Influence of adsorbed water molecules on the volume filling factor}
\label{appendix:water}

It is common practice that the volume filling factor $\phi$ of an agglomerate is determined by measuring the mass $M$ of the agglomerate
\begin{eqnarray}
\phi & = & \frac{M}{\rho V} ,
\end{eqnarray}
where $\rho$ is the density of constituent particles (i.e. $\rho \approx 2.0 \times {10}^3~\mathrm{kg~m^{-3}}$ for amorphous silica).
The number $N$ of particles in the agglomerate is given by 
\begin{eqnarray}
N & = & \frac{3 V \phi}{4 \pi r_0^3} .
\end{eqnarray}
If particles are hydrophilic and adsorb water molecules, then the radius of the particles increase from $r_0$ to $r'_0$ where $\Delta r_0 = r'_0 - r_0$ represents the thickness of water layers.
The adsorption of water molecules reduces the mass of the agglomerate from $M$ to $M'$:
\begin{eqnarray}
M' & = & \frac{4}{3} \pi \left[{{r_0}^3 \rho + \left({{r'_0}^3 - {r_0}^3}\right) \rho_\mathrm{H_2O} }\right] N' ,
\end{eqnarray}
where $\rho_\mathrm{H_2O}$ is the density of water (i.e. $\rho_\mathrm{H_2O} = 1.0 \times {10}^3~\mathrm{kg~m^{-3}}$ and $N'$ is the number of hydrophilic particles encased in the volume $V$ with the filling factor $\phi$:
\begin{eqnarray}
N' & = & \frac{3 V \phi}{4 \pi {r'_0}^3} .
\end{eqnarray}
Accordingly, we have
\begin{eqnarray}
\frac{M'}{\rho V} & = & \left\{{\left({\frac{r_0}{r'_0}}\right)^3 \left[{1 - \left({\frac{\rho_\mathrm{H_2O}}{\rho}}\right)}\right] + \left({\frac{\rho_\mathrm{H_2O}}{\rho}}\right)  }\right\} \phi .
\end{eqnarray}
If the volume filling factor $\phi'$ of an agglomerate is determined by $\phi' = M'(\rho V)^{-1}$ in a laboratory experiment, then the value of the volume filling factor is underestimated, because $M'(\rho V)^{-1} < \phi$.

\section{Surface energy of quartz (crystalline silica)}
\label{appendix:quartz}

\begin{figure}
\center
\includegraphics[angle=-90,scale=0.8]{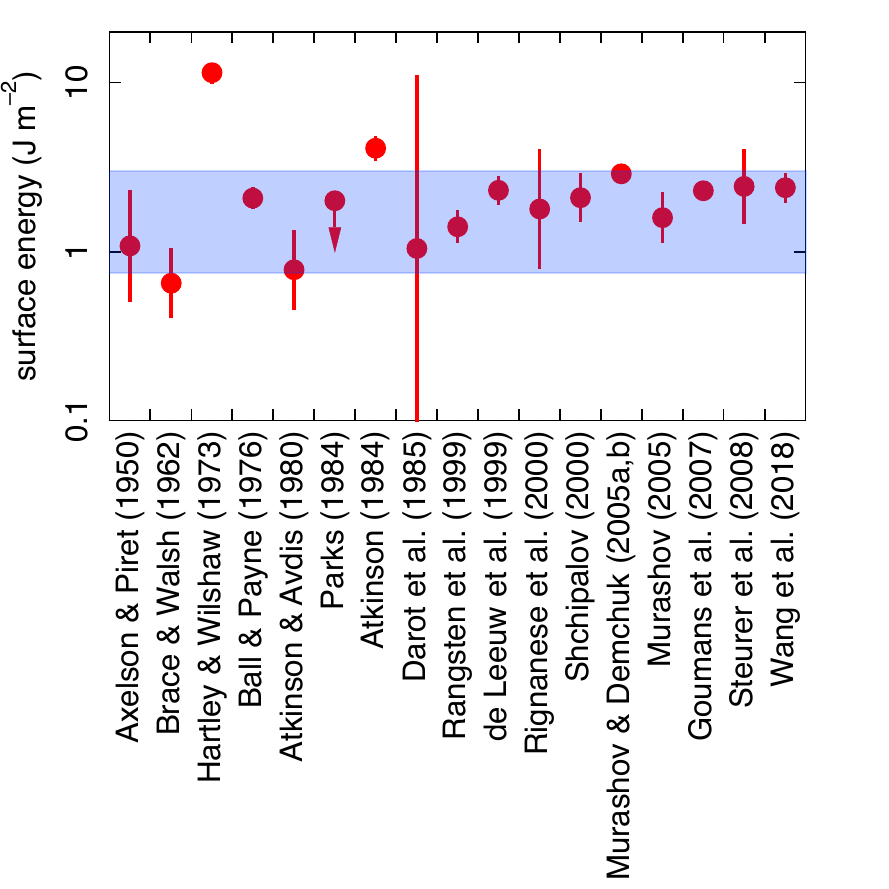}
\caption{The surface energies $\gamma$ of quartz estimated by different methods and authors (the filled circles). Shaded area: $\gamma = 1.5~\mathrm{J~m^{-2}}$ within a factor of 2.
\label{fig:surface-energy-sio2}}
\end{figure}
\citet{axelson-piret1950} listed theoretically evaluated values for the surface energy of quartz in the range of $\gamma = 0.51$--$2.3~\mathrm{J~m}^{-2}$ and took a value of $\gamma = 0.98~\mathrm{J~m}^{-2}$ to investigate their experimental results.
\citet{brace-walsh1962} determined the surface energy of quartz from their measurements by the crack-opening method in the range $\gamma = 0.41$--$1.03~\mathrm{J~m}^{-2}$ depending on its crystallographic axes \citep[see, also][]{tarasevich2006}.
Using the same technique as \citet{brace-walsh1962}, \citet{hartley-wilshaw1973} measured the surface energy of synthetic $\alpha$-quartz to be $\gamma = 11.5 \pm 1.5~\mathrm{J~m}^{-2}$ in air at room temperature.
By using the Vickers hardness test, \citet{atkinson-avdis1980} determined $\gamma = 0.46~\mathrm{J~m}^{-2}$ for quartz (1010) and $\gamma = 1.34~\mathrm{J~m}^{-2}$ for quartz (0001) in air at room temperature, while the surface energy for quartz (1010) was elevated to $\gamma = 1.77~\mathrm{J~m}^{-2}$ at $200^{\circ}\mathrm{C}$. 
\citet{atkinson1984} applied the crack-opening method to measure the surface energy of $\gamma = 3.49$--$4.83~\mathrm{J~m}^{-2}$ for quartz in liquid water or moist air.
\citet{parks1984} argued that the surface energy of quartz is as high as $\gamma = 2~\mathrm{J~m}^{-2}$ in vacuum, by considering effects of adsorbed water molecules on the surface in laboratory experiments.
\citet{darot-et-al1985} applied the Vickers hardness test to estimate the surface energy of $\gamma = 10.75 \pm 0.15~\mathrm{J~m}^{-2}$ for $\alpha$-quartz ($10\bar{1}1$) in argon at room temperature, while they observed a sudden drop of the surface energy down to $\gamma \sim 0.1~\mathrm{J~m}^{-2}$ around the temperature of transition from $\alpha$- to $\beta$-quartz ($10\bar{1}1$), ($10\bar{1}0$) and ($0001$).
\citet{ball-payne1976} estimated the surface energy of quartz to be $\gamma = 1.8$--$2.4~\mathrm{J~m}^{-2}$ using an experimentally derived value of the Si-O bond energy.
\citet{rangsten-et-al1999} derived the surface energy of quartz to be $\gamma = 1.14$--$1.74~\mathrm{J~m}^{-2}$ from their measurements of crack length at elevated temperatures using the crack opening technique.
\citet{deleeuw-et-al1999} and \citet{steurer-et-al2008} computed the surface energy of $\alpha$-quartz (0001) surface to be $\gamma = 1.92$--$2.77~\mathrm{J~m^{-2}}$ and $\gamma = 1.48$--$4.0~\mathrm{J~m^{-2}}$, respectively, using atomistic simulation techniques based on the Born model of ionic solids.
\citet{rignanese-et-al2000} have determined the surface energy for the (0001) surface of $\alpha$-quartz by performing molecular dynamics simulations to be $\gamma = 0.80$--$4.0~\mathrm{J~m}^{-2}$ depending on the model of the surface geometry.
The surface energy for quartz calculated by the periodic density functional theory (DFT) ranges from $\gamma = 2.6$ to $3.2~\mathrm{J~m}^{-2}$ by \citet{murashov-demchuk2005a,murashov-demchuk2005b}, from $\gamma = 1.1$ to $2.2~\mathrm{J~m}^{-2}$ by \citet{murashov2005}, and from $\gamma = 2.2$ to $2.4~\mathrm{J~m}^{-2}$ by \citet{goumans-et-al2007}.
Theoretical calculations by \citet{shchipalov2000} suggest $\gamma =2.875~\mathrm{J~m}^{-2}$ for $\alpha$-cristobalite on the plane \{001\}, $\gamma =1.511~\mathrm{J~m}^{-2}$ for $\alpha$-cristobalite on the plane \{111\}, $\gamma =1.567~\mathrm{J~m}^{-2}$ for $\beta$-cristobalite, and $\gamma =1.707~\mathrm{J~m}^{-2}$ for $\beta$-quartz.
DFT calculations by \citet{wang-et-al2018} gave results that the O-middle termination of quartz (001) surfaces has the lowest surface energy of $\gamma = 1.969~\mathrm{J~m^{-2}}$ and the surface energies of the O-rich termination and the Si termination are $\gamma = 2.892~\mathrm{J~m^{-2}}$ and $\gamma = 2.896~\mathrm{J~m^{-2}}$, respectively.
Figure~\ref{fig:surface-energy-sio2} compiles the surface energies $\gamma$ of quartz estimated by different methods and authors, while most of the values are confined to $\gamma = 1.5~\mathrm{J~m^{-2}}$ within a factor of 2.


\bsp	
\label{lastpage}
\end{document}